\documentclass[preprint2]{aastex}

\slugcomment{Revision 11 (Feb 22 2015; response to referee); to appear in ApJ}

\shorttitle{A Deep X-ray Spectrum of $\delta$ Orionis A\lowercase{a}}
\shortauthors{Corcoran et al.}

\usepackage{natbib}

\usepackage{epstopdf}
\usepackage[usenames]{color}

\usepackage{ulem}

\usepackage{array}

\newcommand{\chandra}{{\it Chandra}}

\newcommand{\asca}{ASCA}
\newcommand{\rosat}{{\it ROSAT}}
\newcommand{\einstein}{{\it EINSTEIN}}

\newcommand{\xmm}{{\it XMM}}

\newcommand{\integral}{\textit{INTEGRAL}}
\newcommand{\most}{{\it MOST}}
\newcommand{\ms}{$M_{\odot}$}

\newcommand{\kms}{km~s$^{-1}$}

\newcommand{\lcgs}{ergs~s$^{-1}$}

\newcommand{\dori}{$\delta$~Ori~A}
\newcommand{\pablo}{\citep{2015arXiv150408002P}}

\begin{document}

\title{A Coordinated X-ray and Optical Campaign on the Nearest Massive Eclipsing Binary, $\delta$~Ori~A\lowercase{a}. I. Overview of the X-ray Spectrum}

\author{M.~F.~Corcoran\altaffilmark{1,2}, J.~S.~Nichols\altaffilmark{3}, H.~Pablo\altaffilmark{4}, T.~Shenar\altaffilmark{5}, A.~M.~T.~Pollock\altaffilmark{6}, W.~L.~Waldron\altaffilmark{7}, A.~F.~J.~Moffat\altaffilmark{4}, N.~D.~Richardson\altaffilmark{4}, C.~M.~P.~Russell\altaffilmark{8}, K.~Hamaguchi\altaffilmark{1,9}, D.~P.~Huenemoerder\altaffilmark{10}, L.~Oskinova\altaffilmark{5}, W.-R.~Hamann\altaffilmark{5}, Y.~Naz\'e\altaffilmark{11, 23}, R.~Ignace\altaffilmark{12}, N.~R.~Evans\altaffilmark{13}, J.~R.~Lomax\altaffilmark{14}, J.~L.~Hoffman\altaffilmark{15}, K.~Gayley\altaffilmark{16}, S.~P.~Owocki\altaffilmark{17}, M.~Leutenegger\altaffilmark{1,9}, T.~R.~Gull\altaffilmark{18}, K.~T.~Hole\altaffilmark{19}, J.~Lauer\altaffilmark{3}, \& R.~C.~Iping\altaffilmark{20,21}}

\altaffiltext{1}{CRESST and X-ray Astrophysics Laboratory, NASA/Goddard Space Flight Center, Greenbelt, MD 20771, USA}
\altaffiltext{2}{Universities Space Research Association, 7178 Columbia Gateway Dr. Columbia, MD 21046, USA}
\altaffiltext{3}{Harvard-Smithsonian Center for Astrophysics, 60 Garden Street, MS 34, Cambridge, MA 02138, USA}
\altaffiltext{4}{D\'epartement de physique and Centre de Recherche en Astrophysique du Qu\'ebec (CRAQ), Universit\'e de Montr\'eal, C.P. 6128, Succ.~Centre-Ville, Montr\'eal, Qu\'ebec, H3C 3J7, Canada}
\altaffiltext{5}{Institut f\"ur Physik und Astronomie, Universit\"at Potsdam, Karl-Liebknecht-Str. 24/25, D-14476 Potsdam, Germany}
\altaffiltext{6}{European Space Agency, \textit{XMM-Newton} Science Operations Centre, European Space Astronomy Centre, Apartado 78, E-28691 Villanueva de la Ca\~{n}ada, Spain}
\altaffiltext{7}{Eureka Scientific, Inc., 2452 Delmer St., Oakland, CA 94602, USA}
\altaffiltext{8}{NASA-GSFC, Code 662, Goddard Space Flight Center, Greenbelt, MD, 20771 USA}
\altaffiltext{9}{Department of Physics, University of Maryland, Baltimore County, 1000 Hilltop Circle, Baltimore, MD 21250, USA}
\altaffiltext{10}{Massachusetts Institute of Technology, Kavli Institute for Astrophysics and Space Research, 77 Massachusetts Avenue, Cambridge, MA 02139 USA}
\altaffiltext{11}{Groupe d'Astrophysique des Hautes Energies, Institut d'Astrophysique et de G\'eophysique, Universit\'e de Li\'ege, 17, All\'ee du 6 Ao\^ut, B5c, B-4000 Sart Tilman, Belgium}
\altaffiltext{12}{Physics and Astronomy, East Tennessee State University, Johnson City, TN 37614, USA.}
\altaffiltext{13}{Harvard-Smithsonian Center for Astrophysics, 60 Garden Street, MS 4, Cambridge, MA 02138, USA}
\altaffiltext{14}{Homer L. Dodge Department of Physics and Astronomy, University of Oklahoma, 440 W Brooks Street, Norman, OK, 73019, USA}
\altaffiltext{15}{Department of Physics and Astronomy, University of Denver, 2112 E. Wesley Avenue, Denver, CO, 80208, USA}
\altaffiltext{16}{Department of Physics and Astronomy, University of Iowa, Iowa City, IA 52242, USA}
\altaffiltext{17}{University of Delaware, Bartol Research Institute, Newark, DE 19716, USA}
\altaffiltext{18}{Laboratory for Extraterrestrial Planets and Stellar Astrophysics, Code 667, NASA/Goddard Space Flight Center, Greenbelt, MD 20771, USA}
\altaffiltext{19}{Department of Physics, Weber State University, 2508 University Circle, Ogden, UT 84408, USA}
\altaffiltext{20}{CRESST and Observational Cosmology Laboratory, NASA/Goddard Space Flight Center, Greenbelt, MD 20771, USA}
\altaffiltext{21}{Department of Astronomy, University of Maryland, 1113 Physical Sciences Complex, College Park, MD 20742-2421, USA}
\altaffiltext{22}{FNRS Research Associate.}

\begin{abstract}
We present an overview of four deep phase-constrained \chandra\  HETGS X-ray observations of \dori. Delta Ori A is actually a triple system which includes the nearest massive eclipsing spectroscopic binary, \dori a, the only such object that can be observed with little phase-smearing with the \chandra\ 
gratings. Since the fainter star, \dori a2, has a much lower X-ray luminosity than the brighter primary (\dori a1), \dori a provides a unique system with which to test the spatial distribution of the X-ray emitting gas around \dori a1 via occultation by the photosphere of, and wind cavity around, the X-ray dark secondary.  Here we discuss the  X-ray  spectrum  
and X-ray line profiles for the combined  observation, having an exposure time of nearly 500 ks and covering nearly the entire binary orbit. The companion papers discuss the X-ray variability seen in the \chandra\ spectra, present new space-based photometry and ground-based radial velocities obtained simultaneous with the X-ray data to better constrain the system parameters, and model the effects of X-rays on the optical and UV spectra. We find that the X-ray emission is dominated by embedded wind shock emission from star Aa1, with little contribution from the tertiary star Ab or the shocked gas produced by the collision of the wind of Aa1 against the surface of Aa2. We find a similar temperature distribution to previous X-ray spectrum analyses.  We also show that the line half-widths are about $0.3-0.5$ times the terminal velocity of the wind of star Aa1. We find a strong  {anti-}correlation between line widths and the line excitation energy, which suggests that longer-wavelength, lower-temperature lines form farther out in the wind.  Our analysis also indicates that the ratio of the intensities of the strong and weak lines  {of \ion{Fe}{17} and \ion{Ne}{10}} are inconsistent with model predictions,  {which may be an effect of resonance scattering}. 

\end{abstract}


\section{Introduction}

Massive O-type stars, though rare, are a primary drivers of the chemical, ionization, and pressure evolution of the interstellar medium. 
The evolution of these stars from the main sequence to supernova depends on their mass and is significantly affected by stellar wind mass-loss. Our best estimates of mass, radius, and luminosity for O stars come from direct dynamical analyses of photometric and radial velocity variations in massive, eclipsing binaries.  However, because massive stars are rare and massive binaries which have been studied in detail rarer still (of the 2386 systems listed in the Ninth Catalog of Spectroscopic Binaries, only 82 of them have O-type components), direct dynamical determinations of stellar parameters are only known for a few systems.    

Current uncertainties regarding the amount and distribution of mass lost through stellar winds are  {even larger, since it is difficult to determine stellar wind parameters in a direct, model-independent way}.  Radiatively driven stellar winds have mass-loss rates of $\dot{M} \sim10^{-5}-10^{-7}$~\ms yr$^{-1}$ \citep[for a review, see][]{2000ARA&A..38..613K}. However, observationally determined mass-loss rates have been estimated, in many, if not most cases, using an idealized smooth, 
spherically symmetric wind.   Stellar winds are probably not spherical; variations of photospheric temperature with latitude are inevitable because of stellar rotation  (and tidal deformation of stars in binaries), and these temperature variations will produce latitudinally dependent wind densities and velocities \citep{1996ApJ...472L.115O}.
Stellar winds are not smooth either; the radiative driving force is inherently unstable to small velocity perturbations,  and wind instabilities are expected to grow into dense structures (clumps) distributed through the wind. In addition,  {clumps} can also be produced by sub-surface convective zones in massive stars caused by opacity peaks associated with the ionization state of helium and iron \citep{Cantiello:2009kq}.  Wind clumps play an important role in determining the overall mass-loss rate, since they carry most of the mass but occupy little volume.   An outstanding question is to determine the number and mass/spatial distribution of  embedded wind clumps. 

%
Collisions  between clumps, or between clumps and ambient wind material at high differential velocities can produce pockets of hot shocked gas embedded in the wind.  Given wind speeds of up to thousands of kilometers per second, these embedded wind shocks  should generate observable X-ray emission \citep[as originally proposed by][]{1980ApJ...241..300L}.  
There have been efforts to determine the fraction of the wind that is clumped, and the radial distribution of the embedded wind shocks, through analysis of the X-ray radiation they produce. 
High spectral resolution X-ray grating spectrometry provides a unique tool to determine the properties of the  {X-ray emitting} hot shocked gas  {produced by embedded wind clumps}. In particular, the forbidden-to-intercombination ($f/i$)  {line} ratios of strong He-like transitions, and analysis of profiles of H-like ions and other strong lines from high resolution spectra (mostly from the \chandra\ and \xmm\ grating spectrometers) indicate that significant X-ray emission exists within 1 to 2 radii of the stellar photosphere  \citep{Waldron:2001lr,2006ApJ...650.1096L,2007ApJ...668..456W}. 
  X-ray lines of strong Ly$\alpha$ transitions (mainly \ion{O}{8}, \ion{Ne}{10}, \ion{Mg}{12}, \ion{Si}{14}, and \ion{S}{16}) show profiles ranging from broad and asymmetric to narrow and symmetric, apparently dependent on stellar spectral type \citep{2009ApJ...703..633W}.  Observed line profile shapes are an important probe of the radius of the maximum X-ray emissivity, modified by absorption from the overlying, cooler, clumped wind. 
 
Clumping-corrected mass-loss rates derived from the analysis of resolved X-ray emission lines \citep{Oskinova:2006dk} are generally in good agreement with predictions of line-driven wind theory, while mass-loss rates derived from analyses of resolved X-ray emission lines are
lower (by a factor of a few) if clumping is not taken into account \citep{Cohen:2014lr}.  
Reducing mass-loss rates by such a large factor would significantly influence our understanding of the ultimate evolution of massive stars.   {However, while i}mportant  wind properties, such as the onset radius of clumping, the fraction of the wind that is clumped, and the radial distribution of clumps through the wind, have been  {indirectly} inferred from detailed X-ray line analysis \citep{Oskinova:2006dk,Owocki:2006fr, Herve:2013rt},
to date, there have been no attempts to determine these properties  {directly.  In this paper, we try to directly constrain the location of the X-ray emitting gas in the wind of a massive eclipsing binary, \dori a, via occultation   by the companion star of the hot gas embedded in the primary's wind}. 

Delta~Ori (Mintaka, HD 36486, 34 Ori) is a visual triple system composed of components A,~B, and C.  Delta~Ori~A itself is composed of a  {massive, short period} close eclipsing system \dori a, and a more distant component, \dori b, which orbits \dori a with a period of 346 years \citep{Tokovinin:2014tg}. The inner binary, $\delta$~Ori~Aa, is the nearest massive eclipsing system in the sky.  It consists of a massive O9.5~II primary (star~Aa1) + a fainter secondary (star~Aa2, B2V-B0.5~III), in a high-inclination ($i>67^{\circ}$), short period ($P=5^{d}.7324$), low eccentricity ($e\approx 0.1$) orbit \citep{Hartmann:1904ly, Stebbins:1915gf,Koch:1981mz, 2002ApJ...565.1216H, 2010A&A...520A..89M}. Because it is nearby, bright, with a high orbital inclination, \dori a is an important system since it can serve as a fundamental calibrator of the mass-radius-luminosity relation in the upper HR diagram. It is disconcerting, though, that published stellar masses for the primary star \dori a1 are different by about a factor of two 
\citep{2002ApJ...565.1216H, 2010A&A...520A..89M}\footnote{Some progress has been recently made by \cite{Harmanec:2013lr} and by \cite{Richardson:2015fk} in disentangling lines of \dori a2 from \dori a1 and \dori b in the composite spectrum}.

Delta Ori Aa is also a bright X-ray source \citep{Long:1980dp, Snow:1981tg, Cassinelli:1983qe} and is the only eclipsing short-period O-type binary system that is bright enough to be observable with the \chandra\ gratings with little phase smearing, offering the chance to study of variations of the X-ray emission line profiles as a function of the orbital phase.

Since the luminosity of the secondary, \dori a2, is less than 10\% that of the primary, and since X-ray luminosity scales with stellar bolometric luminosity \citep{Pallavicini:1981ai,Chlebowski:1989cr, 1997A&A...322..167B} for stars in this mass range, it should also be less than 10\% as bright in X-rays as the primary.  Thus the X-ray emission from the system is dominated by the hot gas in the wind of the primary star. Therefore, occultation of different X-ray-emitting regions in the   wind  of \dori a1 by the photosphere and/or wind of the  {X-ray faint} secondary, \dori a2, presents the opportunity  to directly study the radial  distribution of the hot shocked gas in the primary's wind,  by measuring occultation effects in X-ray line emission as a function of ionization potential and orbital phase.
Since X-ray lines of different ionization potentials are believed to form at different   radial distances above the primary's surface, differential variations in the observed set of X-ray lines as a function of orbital phase  {allow} us to probe the hot gas distribution within the primary wind's acceleration  {zone,} where most of the X-ray emission is believed to originate.  He-like ions in the X-ray spectrum provide a complementary measure of the radial distribution of the hot gas, since these lines are sensitive to wind density and the dilute ambient UV field. This makes \dori a a unique system with which to 
constrain directly the spatial distribution of X-ray emitting clumps embedded in the wind of an important O star.  The main challenge, however, is the relatively small size of \dori a2 compared to the  {size of the} X-ray emitting region, since the hot gas is  {expected to be} distributed in a large volume throughout the stellar wind.

This paper provides an overview of the X-ray grating spectra obtained during a 479
ksec \chandra\   {campaign on} \dori a+Ab in 2012.  The purpose of this project was to obtain high signal-to-noise observations with the \chandra\ High Energy Transmission Grating Spectrometer \citep[HETGS;][]{Canizares:2005bh} of \dori a over almost an entire binary orbit, including key orbital phases, with coordinated ground-based radial velocity monitoring at H$\alpha$ and \ion{He}{1}~6678 (primarily obtained by a group of amateur astronomers), and high precision, simultaneous photometry from space by the Canadian Space Agency's Microvariability and Oscillations of Stars telescope  {\citep[\most, ][]{Walker:2003yq}}.  This paper provides an overview of the combined HETGS spectrum from our four observations, and is organized as follows.  In Section \ref{sec:obs} we present a summary of the four observations and discuss the acquisition and reduction of the data sets.  Section \ref{sec:image} presents an analysis of the zeroth-order image of the system to constrain the X-ray contribution of \dori b to the observed X-ray emission. Section \ref{sec:combined} presents the temperature distribution and overall properties of the strong emission lines in the combined spectrum of the four observations. Section \ref{sec:cw} discusses the possible influence of the collision of the wind from the primary with the weak wind or photosphere of the secondary, and the influence of any such collision on the wind's thermal and density structure.
We present conclusions in Section~\ref{sec:conc}. A series of companion papers  presents the results of the variability analysis of the X-ray continuum and line emission \cite[][in press, Paper II]{2015arXiv150704972N}, the ground-based radial velocity and \most\ space-based photometric monitoring and analysis \cite[][in press, Paper III]{2015arXiv150408002P}, and a complete non-LTE analysis of the spectral energy distribution of \dori a+b from optical through X-rays \cite[][in press, Paper IV]{2015arXiv150303476S}.

\section{Stellar And System Parameters}
\label{sec:params}
The stellar parameters given by \cite{2002ApJ...565.1216H} and \cite{2010A&A...520A..89M} differ significantly, and this difference has important consequences for our understanding of the evolutionary state of the system, and the influence of mass-loss and/or non-conservative mass transfer. 
\citet{2002ApJ...565.1216H} derived masses of $M_{Aa1} = 11.2M_{\odot}$ and $M_{Aa2}=5.6M_{\odot}$ for the primary and secondary stars, making the primary significantly overluminous for its mass (or undermassive for its spectral type).  The radial velocity and photometric analysis of \cite{2010A&A...520A..89M} were consistent with a substantially higher mass for the primary, $M_{Aa1}=25M_{\odot}$, after a correction for perceived contamination of the radial velocity curve by lines from \dori b.   
Whether the O9.5~II primary has a normal mass and radius for its spectral type is important for understanding the history of mass exchange/mass-loss from \dori a, and how this history is related to the current state of the radiatively driven wind from the primary.  

An important goal of our campaign is to derive definitive stellar and system parameters for \dori a. To this end, we obtained high-precision photometry of the star with the \most\ satellite, along with coordinated ground-based optical spectra to allow us to obtain contemporaneous light- and radial-velocity curve solutions, and to disentangle the contributions from Aa2 and/or Ab from the stellar spectrum.  We also performed an analysis of the optical and archival IUE UV spectra using the non-LTE Potsdam Wolf-Rayet (PoWR) code \citep{Grafener:2002vn,Hamann:2003rt}. The light curve and radial velocity curve analysis is presented in Pablo et al. (2015), while the non-LTE spectral analysis is presented in Shenar et al. (2015).  Table \ref{tab:params} summarizes these results.  In this table, the values and errors on the parameters derived from the \most\ photometry and radial velocities are given for the low-mass solution provided in Pablo et al. (2015). Note that we find better agreement between the derived stellar parameters (luminosities, masses, radii, and temperatures) and the spectral type of \dori a1 if we use the $\sigma$-Orionis cluster distance \citep[$d = 380$~pc,][]{Caballero:2008oq} for \dori, rather than the smaller Hipparcos distance. 
Therefore, we adopt $D=380$~pc as the distance to \dori\ (for a full discussion of the distance to \dori, see Shenar et al. 2015). The spectral type of \dori a2 is not well constrained; \cite{2002ApJ...565.1216H} assign it a spectral type of B0.5 III, while \cite{2010A&A...520A..89M} do not assign a spectral type due to the difficulty in identifying lines from the star.  Shenar et al. (2015) assign an early-B dwarf spectral type to \dori a2 ($\approx$ B1V).

{\setlength{\extrarowheight}{3pt}
\begin{table*}[htp]
\caption{Stellar, Wind and System parameters for \dori a1+Aa2 from Analysis of the Optical, UV and X-ray spectra \citep{2015arXiv150303476S} and the Solution to the MOST Light Curve and Ground-Based Radial Velocities \citep{2015arXiv150408002P}.}
\begin{center}
\begin{tabular}{|l| >{\centering\arraybackslash}m{2in} | >{\centering\arraybackslash}m{2in} |}
\cline{1-3}
& \multicolumn{2}{c|}{Method} \\
\cline{2-3}
\multicolumn{1}{|c|}{Parameter}& \multicolumn{1}{c|}{POWR Analysis$^{a}$} & \multicolumn{1}{c|}{light curve \& RV solution$^{b}$} \\
\hline
 $T_{\rm eff} $[kK] (Aa1) &  $29.5\pm 0.5$  & 30 (adopted)  \\
 $T_{\rm eff} $[kK] (Aa2) &  $25.6\pm 3$&$24.1^{+0.4}_{-0.7}$\\
 $ R [R_\odot] $ (Aa1) &  $16.5\pm 1$ & $15.1$    \\
 $ R [R_\odot] $ (Aa2) &  $6.5^{+2}_{-1.5}$  & $5.0$     \\
 $ M [M_\odot]$  (Aa1) &  $24^{+10}_{-8}$  & $23.8$   \\
 $ M [M_\odot]$  (Aa2) &  $8.4^{e}$  &  $8.5$       \\ 
 $ L\, [\log L_\odot] $  (Aa1) &  $5.28\pm 0.05$  & 5.20
\\
 $ L\, [\log L_\odot] $  (Aa2) &  $4.2\pm 0.2$  & 3.85
\\
\hline 
 $ v_\infty $ [\kms] (Aa1) &  $2000\pm 100$  &    \\
 $ v_\infty $ [\kms] (Aa2) &  $1200^{e}$  &    \\
 $\log \dot{M}\,[M_\odot /{\rm yr}] $ (Aa1) &  $-6.4 \pm 0.15$  &    \\
 $\log \dot{M}\,[M_\odot /{\rm yr}] $ (Aa2) & $\le -6.8$  &    \\
 $ E_{B-V} $ (ISM) &  \multicolumn{1}{c|}{$0.065\pm 0.002$}  & \\
 $ A_{V} $ (ISM)   &  \multicolumn{1}{c|}{$0.201\pm 0.006$}  & \\
 $ \log N_{\rm H} $ (ISM)  &  \multicolumn{1}{c|}{$20.65\pm 0.05$} & \\
\hline
 $ P [d] $ & & $ 5.732436^{d}$    \\
 $E_{0}$ (primary min, HJD) &  &  $2456277.790\pm0.024$     \\ 
 $T_{0}$ (periastron, HJD) & & $2456295.674\pm0.062$ \\
 $ a [R_\odot] $  & & $ 43.1
\pm 1.7$      \\
 $ i $ [deg.] & & $ 76.5
\pm 0.2
$    \\
 $\omega$ [deg.]& & $141.3
\pm 0.2$  \\
 $\dot{\omega}$ [deg. yr$^{-1}$] & & $ 1.45 \pm 0.04 $ \\
 $ e $ & & $0.1133
\pm 0.0003$    \\
 $\gamma$~[\kms] & &  $15.5
\pm 0.7$ \\
 \hline
 Sp. Type (Aa1) & \multicolumn{2}{c|}{O9.5II$^{a,c,d}$}  \\
 Sp. Type (Aa2) & \multicolumn{2}{c|}{B1V$^{a}$} \\
 \hline
 $ D $ [pc]  &  \multicolumn{2}{c|}{$380$ (adopted)}  \\

\hline
\multicolumn{3}{l}{Notes:} \\
\multicolumn{3}{l}{\footnotesize{$^{a}$Shenar et al. (2015); $^{b}$from the low-mass model solution of Pablo et al. (2015);}} \\
\multicolumn{3}{l}{\footnotesize{$^{c}$\cite{Maiz-Apellaniz:2013uq}; $^{d}$Mayer et al. (2010); $^{e}$Adopted assuming a spectral type of B1V.}} \\
\end{tabular}
\end{center}
\label{tab:params}
\end{table*}
\section{Previous X-ray Observations}

X-ray emission from $\delta$~Ori was first tentatively identified via sounding rocket observations \citep{Fisher:1964ij}.  X-ray imaging spectrometry of \dori\ at low or modest resolution was obtained by the \einstein\ \citep{Long:1980dp}, 
\rosat\ \citep{1993A&A...280..519H}, and \asca\
\citep{1994ApJ...436L..95C} X-ray observatories.  Its X-ray luminosity is typically  $L_{x}\sim10^{31-32}$~\lcgs, with $L_{x}/L_{bol}\approx10^{-7}$ in accord with the canonical relation for massive stars \citep{Pallavicini:1981ai,Chlebowski:1989cr,1997A&A...322..167B}. The X-ray spectrum of \dori\ was  observed at high resolution by X-ray grating spectrometers on \chandra\ in two previous observations at restricted orbital phases.  An analysis of a fifty kilosecond \chandra\ HETGS spectrum from 2000 January 13 by \cite{2002ApJ...577..951M} revealed strong line emission from O, Ne, Mg, and Fe, along with weaker emission from higher-ionization lines like Si XIII and S XV, and unusually narrow line half-widths of $\approx400$~\kms. Using a simple analysis taking into account dilution of the photospheric UV field and a $1/r^{2}$ falloff in wind density, \cite{2002ApJ...577..951M} derived formation regions for the dominant He-like ions \ion{Mg}{11}, \ion{Ne}{9}, and 
\ion{O}{7} extending just above the stellar photosphere to 3--10 times the photospheric radius.  An analysis of a 100 ks \chandra\ Low Energy Transmission Grating Spectrometer \citep[LETGS;][]{Brinkman:1987rf} + High Resolution Camera observation from 2007 November 09 by \cite{Raassen:2013fk} also showed that the \ion{Mg}{11}, \ion{Ne}{9}, and \ion{O}{7} emission regions extend from 2--10 stellar radius, and showed that the longer wavelength ions like \ion{N}{6} and \ion{C}{5} form at substantially greater distances from the star (50--75 times the stellar radius), and that the spectrum could be modeled by a three-temperature plasma in collisional ionization equilibrium with temperatures of 0.1, 0.2, and 0.6 keV.

{\setlength{\extrarowheight}{2pt}
\begin{table*}[htp]
\caption{New \chandra\ Observations of \dori a+Ab}
\begin{center}
\begin{tabular}{lllllllllr}
\hline
ObsID & Start  & Start  & End  & End     &  Midpoint & Midpoint & $\Delta T$ &  Exposure & Roll \\
      & HJD    & Phase  & HJD  & Phase   &   HJD     & Phase    &  Days      &  s        & deg.  \\
\hline 
14567 & 2456281.21 & 396.604 & 2456282.58 & 396.843 & 2456281.90 & 396.724 & 1.37 & 114982 & 345.2 \\
14569 & 2456283.76 & 397.049 & 2456285.18 & 397.297 & 2456284.47 & 397.173 & 1.42 & 119274 & 343.2 \\
14570 & 2456286.06 & 397.450 & 2456287.52 & 397.705 & 2456286.79 & 397.578 & 1.46 & 122483 &  83.0 \\
14568 & 2456288.67 & 397.905 & 2456290.12 & 398.159 & 2456289.39 & 398.032 & 1.45 & 121988 & 332.7 \\
\hline
\end{tabular}
\end{center}
\label{tab:obs}
\end{table*}
\section{New \chandra\ Observations}
\label{sec:obs}

A listing of the \chandra\ observations of \dori a+Ab obtained as part of this campaign is given in Table \ref{tab:obs}. These observations were obtained with the \chandra\  HETGS+ACIS-S spectrometric array.  The HETGS consists of 2 sets of gratings: the Medium Resolution Grating (MEG) covering the range 2.5--26\AA\ and and the High Resolution Grating (HEG) covering the range 1.2--15\AA; the HEG and MEG have resolving powers of $\lambda/\Delta\lambda\approx 1000$ at long wavelengths, falling to $\sim100$ near 1.5\AA\ \citep{Canizares:2005bh}.
Four observations covering most of the orbit were obtained within a 9-day timespan to reduce any influence of orbit-to-orbit X-ray variations, for a combined exposure time of 479~ks. Table \ref{tab:obs} lists the start and stop HJD, phases, and exposure durations for the four individual observations. Figure \ref{fig:obs} shows the time intervals of each observation superposed on the simultaneous MOST optical light curve of \dori\ \pablo. The Chandra observations provide both MEG and HEG dispersed first order spectra as well as the zeroth order image. Due to spacecraft power considerations as well as background count rate issues, it was necessary to use only five ACIS CCD chips instead of six; thus, chip S5 was not used. This means that wavelengths longer than about 19~\AA\ in the MEG plus-side dispersed spectrum and about 9.5~\AA\ in the HEG plus-side dispersed spectrum are not available.  Therefore, the strong \ion{O}{7} line at 21~\AA\ was only observed in the MEG-1 order. The buildup of contaminants on the ACIS-S optical blocking filters with time further degraded the long wavelength  {sensitivity} for all  {first-order spectra}. Each of the four observations  {experienced} a large variation in focal plane temperature during the observation. While a temperature-dependent calibration is applied to each observation in standard data processing, the calibration is based on a single temperature measurement taken at the end of the observation. In particular, the focal plane temperature for portions of each observation exceeded the temperature at which the temperature-dependent effects of charge transfer inefficiency (CTI) are calibrated \citep{Grant:2006kq}. This could cause residual errors in the correction of pulse heights for those portions of the observations in the high-temperature regime. 

Each ObsID was processed using the standard processing pipeline used in production of the  \chandra\ Transmission Grating Data Archive and Catalog \citep[TGCAT;][]{Huenemoerder:2011vn}.  Briefly, event filtering, event transformation, spectral extraction, and response generation are done with standard \chandra\ Interactive Analysis of Observations software tools \citep{Fruscione:2006uq} as described in detail by \cite{Huenemoerder:2011vn}.  This pipeline produces standard X-ray events, spectra, responses, effective areas, aspect histograms, and light curves.  
We used version 4.5.5 of the Chandra Calibration Database (CALDB), along with CIAO version 4.5 \& 4.6 in the analysis presented here.  
In order to examine  variability, the data were also divided into  $\sim10$~ks segments, and spectra, response files, effective areas and light curves were generated for each segment. Analysis of the time-sliced data is presented in \cite{2015arXiv150704972N}.  

\begin{figure*}[htbp]    \centering
      \includegraphics[width=6in]{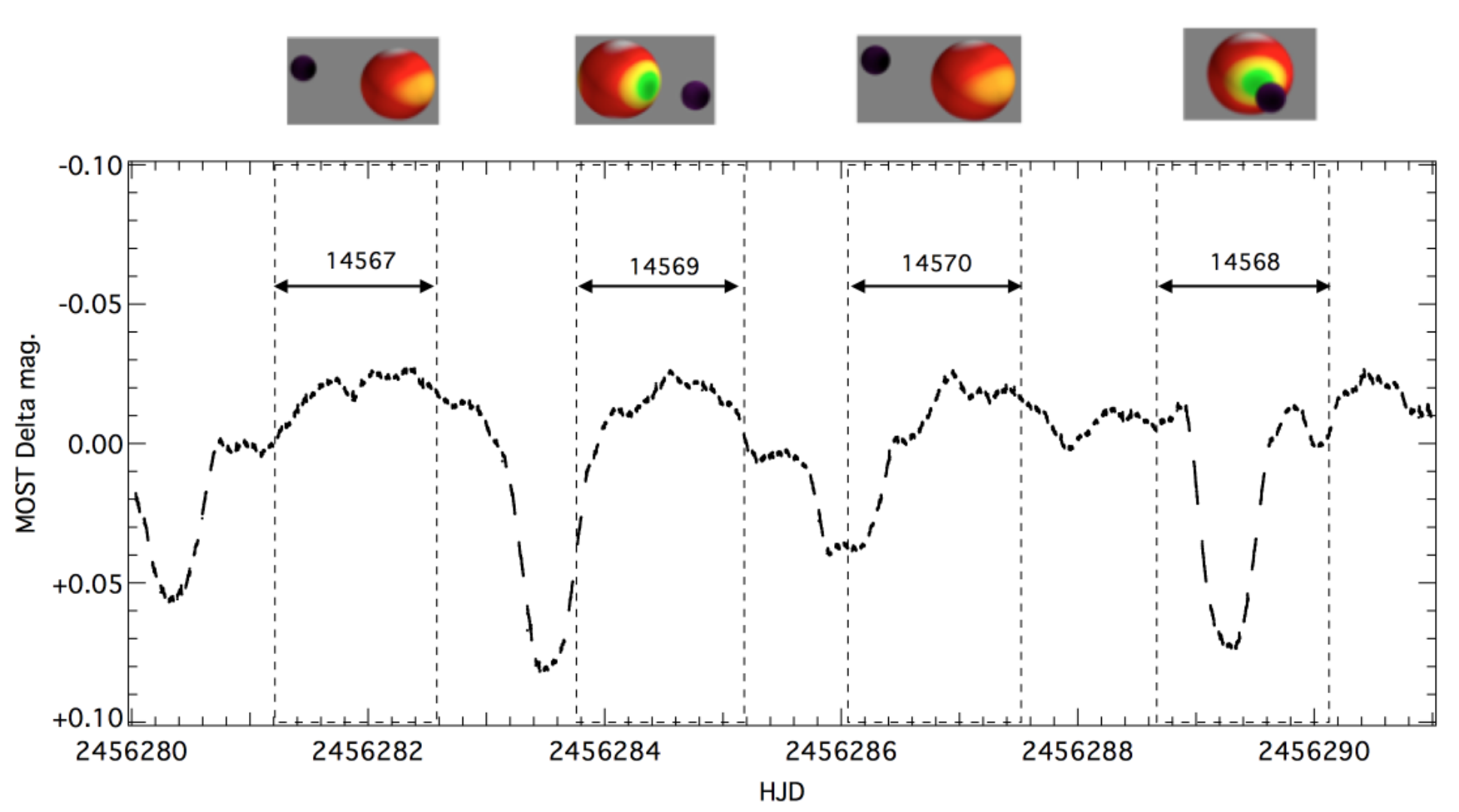}
   \caption{Timings of the \chandra\ observations along with the \most\ light curve.  The images above the plot show the orientations of \dori a1 and Aa2 near the midpoint of the observation according to the photometric and spectroscopic analysis of Pablo et al. (2015). In the images, the orbital angular momentum vector lies close to the plane of the paper and points to the top of the page.}
   \label{fig:obs}
\end{figure*}

\newpage

\section{Analysis of the X-ray Image}
\label{sec:image}

The \dori a1,2 inner binary is orbited by a more distant tertiary component (\dori b) at a current projected separation of $0''.3$  with an orbital period of $\approx 346$ years \citep{Tokovinin:2014tg}. This separation is just below the spatial resolution of \chandra, and thus \chandra\ imaging observations allow us to spatially examine 
the X-ray contribution from the Ab component.  Figure~\ref{fig:acis} shows unbinned zeroth-order images from our four HETGS+ACIS  observations, along with the expected location of Ab and the Aa pair at the times of the \chandra\ observations in 2012.  

To constrain the X-ray contribution of \dori b, we generated zeroth-order images for the four individual pointings 
listed in Table~\ref{tab:obs}, using the Energy-Dependent Subpixel Event Repositioning (EDSER\footnote[23]{\url{http://cxc.harvard.edu/ciao4.4/why/acissubpix.html}}) method to generate images with a pixel size of $0''.125$.  
We generated images in 0.3--1 and 1--3 keV bands,  but found no significant 
differences in any of the four observations when we compared the soft and hard band images.  For each image, we then applied the CIAO tool SRCEXTENT to calculate the size and associated uncertainty of the photon-count source image or using the Mexican Hat Optimization  algorithm\footnote[24]{http://cxc.harvard.edu/ciao/ahelp/srcextent.html
}. 

The results of the SRCEXTENT analysis are given in Table \ref{tab:srcextent}.  The derived major and minor axes of each image are equal and consistent with the \chandra\ point spread function, $\sim0.3''$.  The peak of the image is consistent with the location of the Aa component, and is about a factor of two farther than the Ab component.  We conclude that the peak positions of the zeroth-order images indicate that Aa is the primary X-ray source, with little or no contribution from Ab.
Our analysis also suggests that the ObsID 14568 image may be slightly elongated, which may indicate a possible issue with the instrumental pointing or aspect reconstruction for this observation.

{\setlength{\extrarowheight}{3pt}
\begin{table*}[htp]
\caption{SRCEXTENT Analysis Results}
\begin{center}
\begin{tabular}{l|lccrcc}
\hline
      & Band & Major Axis & Minor Axis & PA & Peak distance Aa & Peak distance Ab \\
ObsID & keV & arcsec & arcsec & deg. & arcsec & arcsec \\
\hline
14567 & 0.3--1 & 0.34 & 0.33 & 83.3 & 0.19 & 0.40 \\
 & 1--3 & 0.32 & 0.28 & 83.8 & 0.19 & 0.42 \\ \hline
14569 & 0.3--1 & 0.32 & 0.32 & 32.1 & 0.23 & 0.44 \\
 & 1--3 & 0.29 & 0.28 & 27.6 & 0.25 & 0.47 \\ \hline
14570 & 0.3--1 & 0.32 & 0.32 & 136.9 & 0.09 & 0.35 \\
 & 1--3 & 0.26 & 0.22 & 48.3 & 0.08 & 0.34 \\ \hline
14568 & 0.3--1 & 0.51 & 0.32 & 35.9 & 0.24 & 0.41 \\
 & 1--3 & 0.48 & 0.25 & 31.2 & 0.24 & 0.42 \\
\hline
\end{tabular}
\end{center}
\label{tab:srcextent}
\end{table*}
\begin{figure*}[htbp]    \centering
      \includegraphics[width=6in]{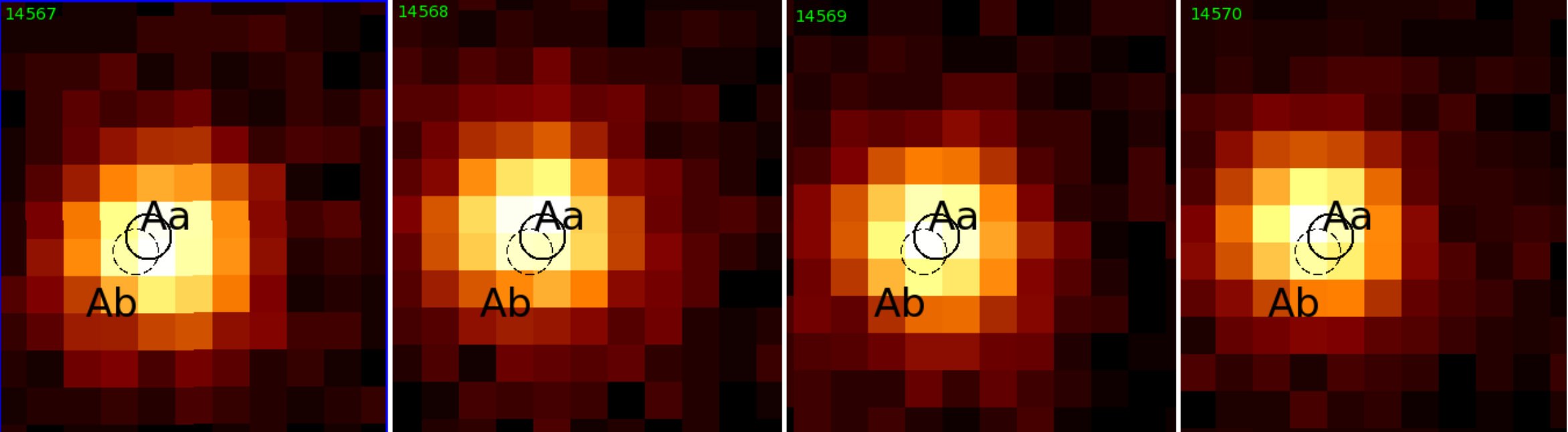}   
      \caption{
Unbinned images from the four ObsIDs listed in Table \ref{tab:obs}. ObsIDs, left to right: 14567,  14568, 14569, and 14570. 
The positions of Aa and Ab are shown by the full and dashed circles, respectively.}
\label{fig:acis}
\end{figure*}

\section{Combined Spectrum}
\label{sec:combined}

Figure \ref{fig:specid} shows the co-added spectrum from the four  observations, with a total exposure of 479~ks. This represents the second longest exposure yet obtained on a massive star at wavelengths $\lesssim8$~\AA\ and a resolving power of $\lambda/\Delta\lambda>400$. 
The strongest lines are \ion{O}{8}, \ion{Fe}{17}, \ion{Ne}{9} \& \ion{Ne}{10}, \ion{Mg}{11} \& \ion{Mg}{12}, and \ion{Si}{13}. 

\begin{figure*}[htbp]    \centering
     \includegraphics[width=5in]{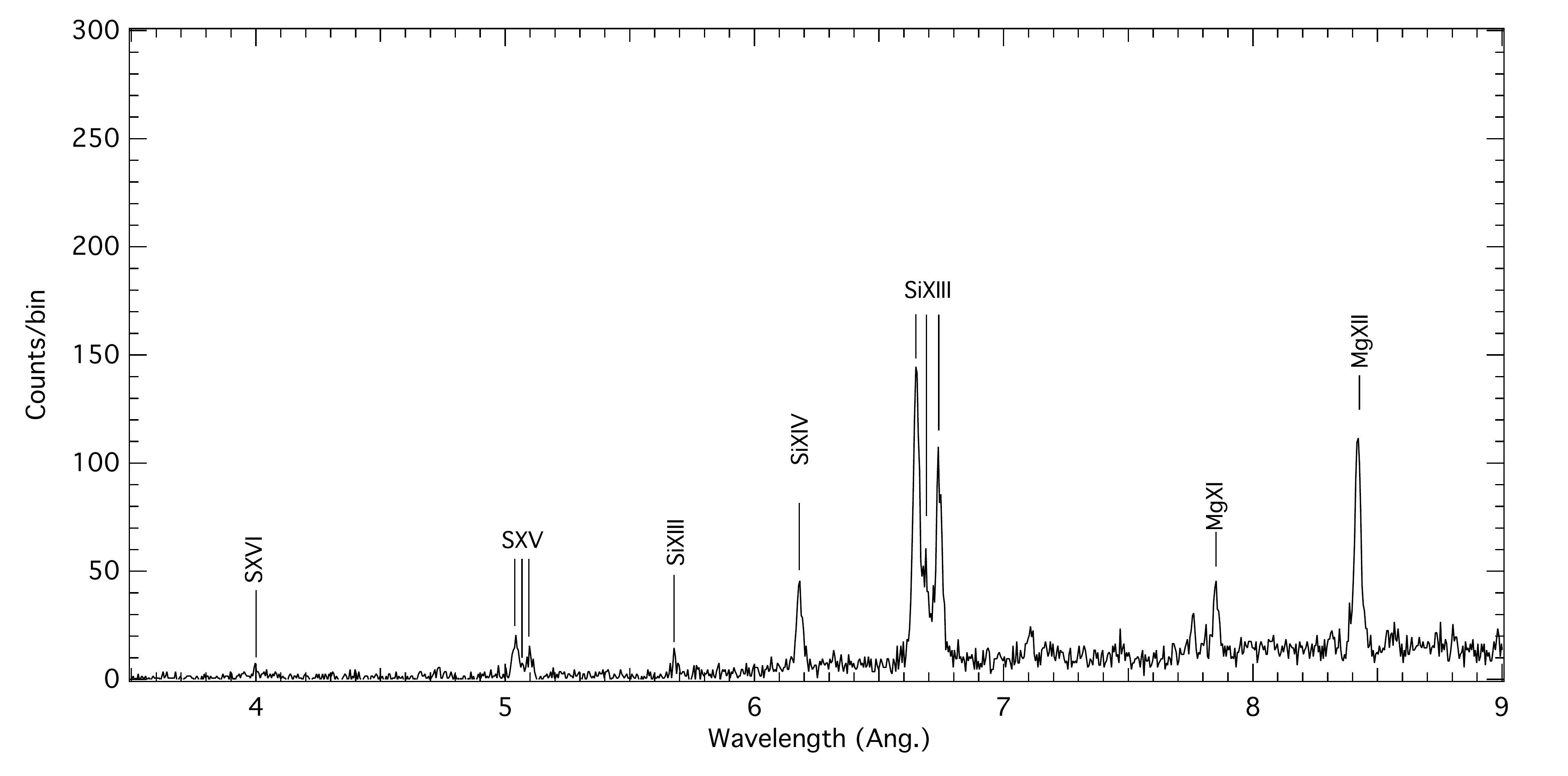}
   \includegraphics[width=5in]{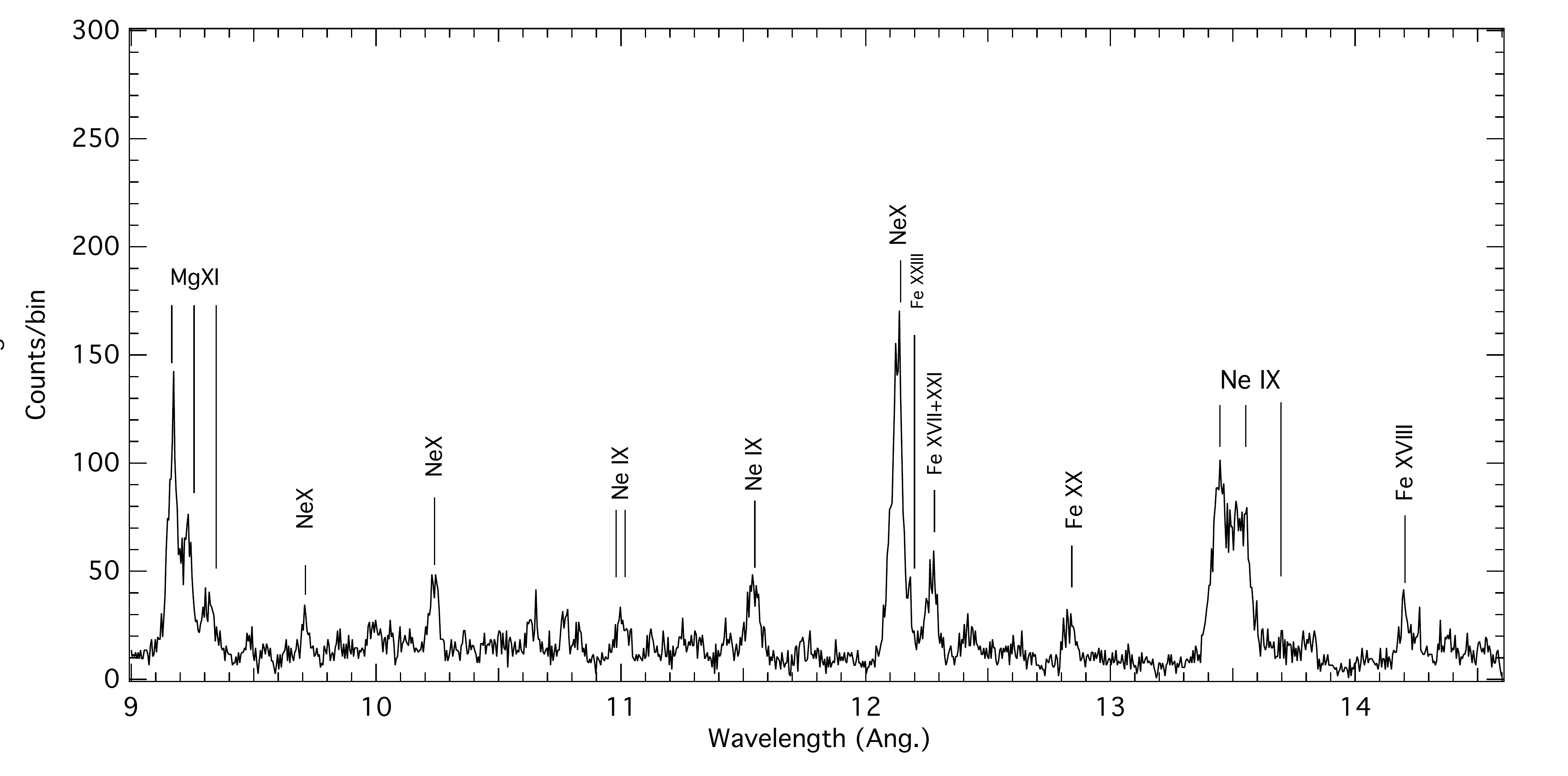}
   \includegraphics[width=5in]{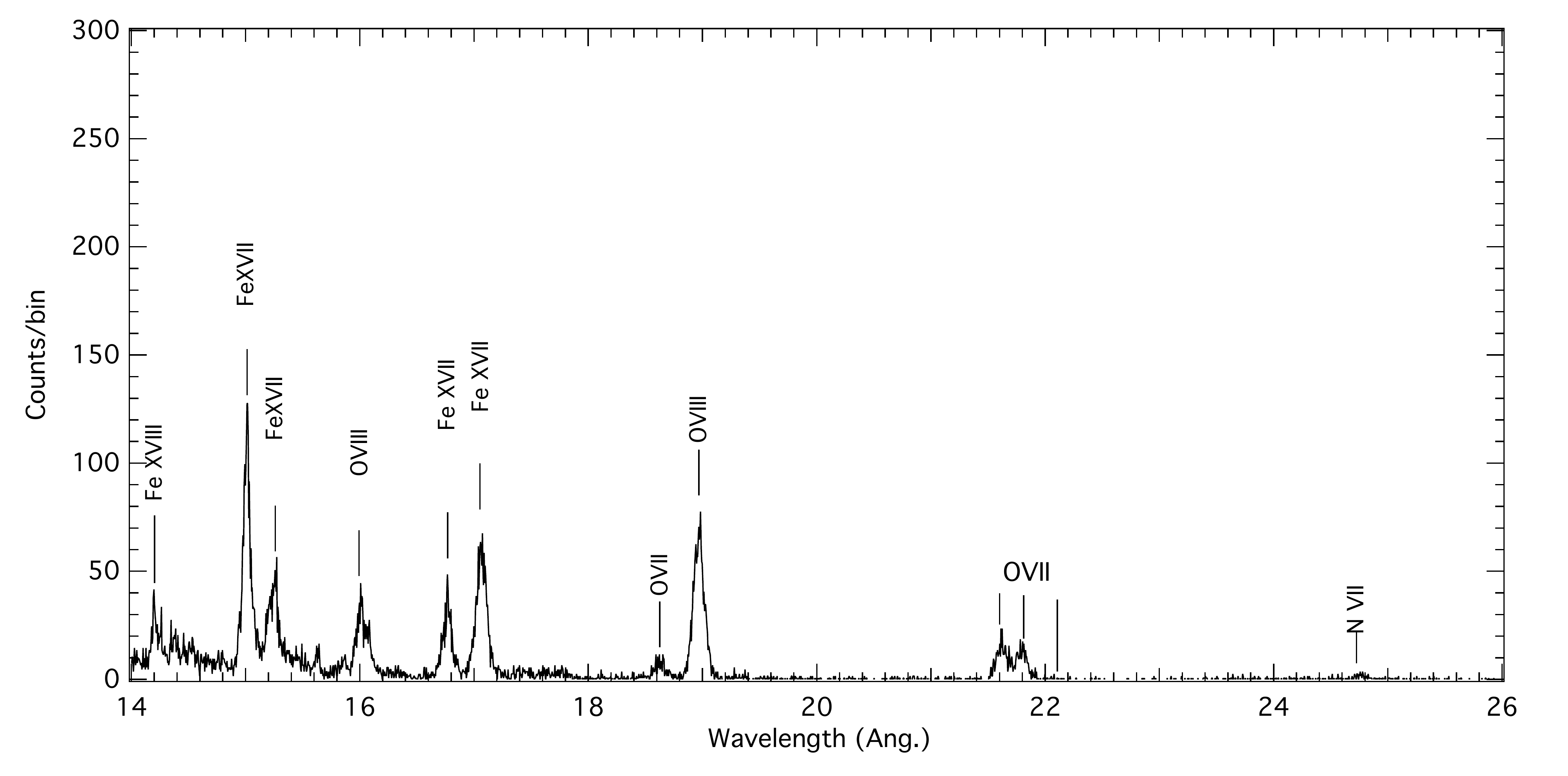}

   \caption{The combined  MEG+HEG spectrum of \dori, from 3.5~\AA\ to 26~\AA.}
   \label{fig:specid}
\end{figure*}

\newpage

\subsection{Temperature Distribution}
\label{sec:temps}

We modeled the combined spectrum with a combination of absorbed collisional ionization equilibrium models using 
the Interactive Spectral Interpretation System \citep[ISIS;][]{Houck:2000qd}. The model we applied includes two low-temperature components seen through a common absorption component, plus a third hotter component with its own absorption component to account for any contribution from a hot colliding wind region embedded within the wind of the binary (see Section \ref{sec:cw} below).  In ISIS terminology, the mode we used was ``(xaped(1) + xaped(2)) * TBabs(3) + xaped(4) * TBabs(5)'', where ``xaped'' represents emission from an optically thin plasma in collisional ionization equilibrium based on the ATOMDB atomic database version 2.0.2 \citep{Smith:2000qv, Foster:2012sf}, and ``TBabs'' represents interstellar absorption 
%
%
\citep{Wilms:2000rr}. Solar abundances were assumed for both the emission and absorption components\footnote[25]{\citet{2015arXiv150303476S} show that N and Si are slightly sub-solar, but these differences are not significant for our analysis.}. This model is an approximation to the actual temperature distribution and absorption, but is the simplest one we found that adequately describes the observed grating spectrum. We allowed for velocity broadening of the emission lines, with turbulent velocity broadening constrained to be less than roughly twice the maximum wind terminal velocity, 3000~\kms. We allowed the line centroid velocities of the three emission components to vary, but found that overall the line centroids are unshifted in the combined spectrum. Figure \ref{fig:specfit} compares the best-fit model to the data, while the model components are given in Table \ref{tab:specfit}. In this table, we also convert the derived turbulent velocities $V_{turb}$ to equivalent line half-widths at half maximum, using \ion{O}{8}, \ion{Ne}{10} and \ion{Mg}{12} for the low-, medium-, and high-temperature components, respectively.

{\setlength{\extrarowheight}{3pt}
\begin{table*}[htp]
\caption{Best-Fit to the Combined HETGS spectrum. The adopted model is (APED$_{1}$+APED$_{2}$)*$N_{H,1}$+APED$_{3}$*$N_{H,2}$}
\begin{center}
\begin{tabular}{llr}
\hline
Component & Parameter & Value \\
\hline
& $T_{1}$ (MK) & 1.25 \\
1 &$EM_{1}$ ($10^{55}$ cm$^{-3}$) & 4.46 \\
& $V_{turb,1}$ (\kms) & 1313\\
& HWHM (\kms) & 1094
\\ \hline
& $T_{2}$ (MK) & 3.33 \\
2 & $EM_{2}$ ($10^{55}$ cm$^{-3}$)  & 0.87 \\
& $V_{turb,2}$ (\kms) & 1143 \\
& HWHM (\kms) & 953
\\ \hline
Absorption 1 & $N_{H,1}$ ($10^{22}$ cm$^{-2}$) & 0.14 \\
\hline
& $T_{3}$ (MK) & 9.11 \\
3  & $EM_{3}$ ($10^{55}$ cm$^{-3}$) & 0.26 \\
& $V_{turb,3}$ (\kms) & 685 \\
& HWHM (\kms) & 574
\\ \hline
Absorption 2 & $N_{H,2}$  ($10^{22}$ cm$^{-2}$)& 0.24 \\
\hline
$f_{x}$ (ergs cm$^{-2}$ s$^{-1}$) (observed, $1.7-25$~\AA) & & $8.2\times10^{-12}$ \\
$L_{x}$ (ergs s$^{-1}$) (observed, $1.7-25$~\AA) & & $1.4
\times 10^{32}$ \\
$\log L_{x}/L_{bol}$ & & $-6.73$ \\
\hline
EM-weighted Average Temperature (MK) & \multicolumn{2}{r}{1.94} \\
\hline
\end{tabular}
\end{center}
\label{tab:specfit}
\end{table*}
\begin{figure*}[htbp]    \centering
      \includegraphics[width=5in]{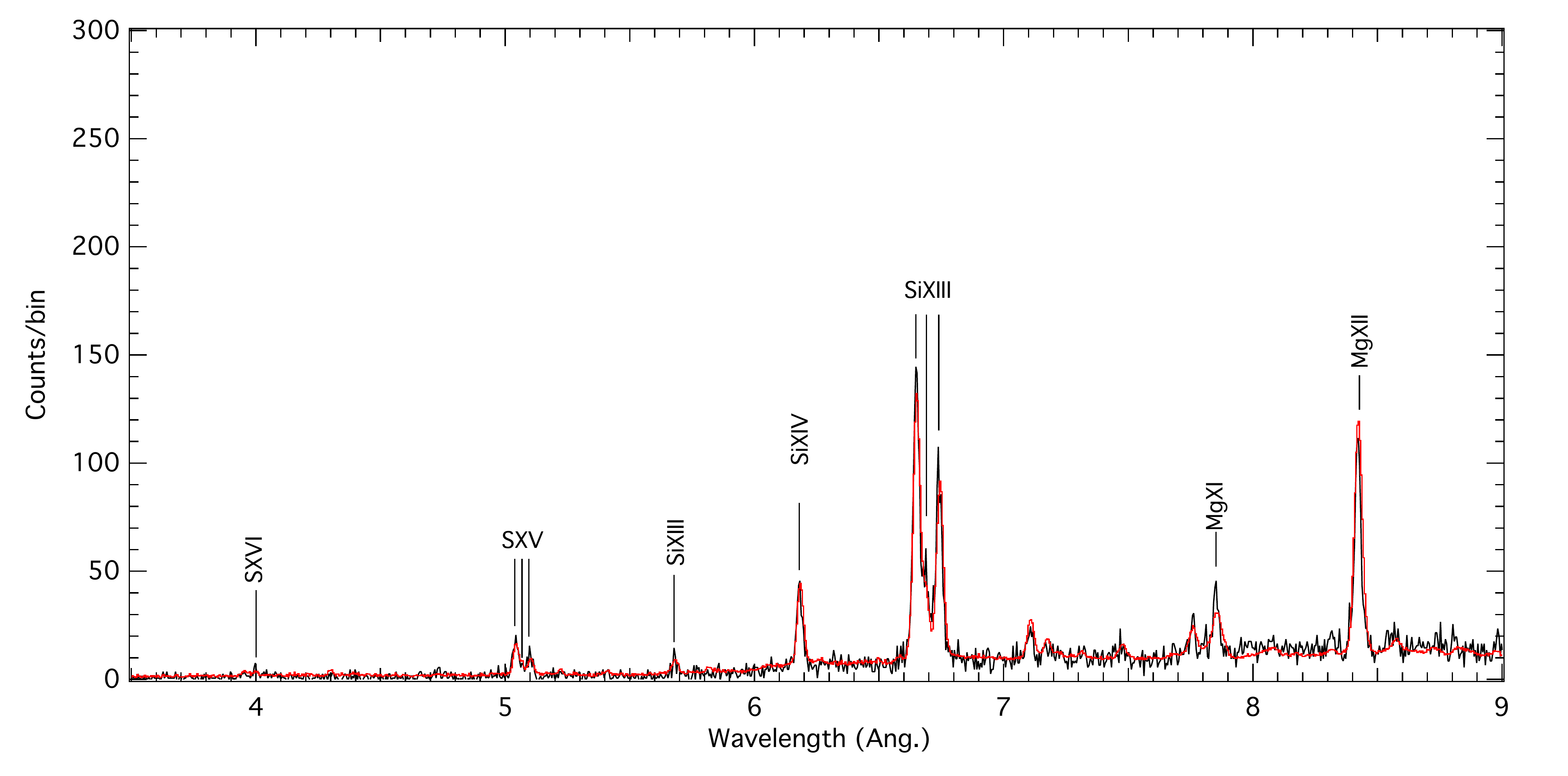}
   \includegraphics[width=5in]{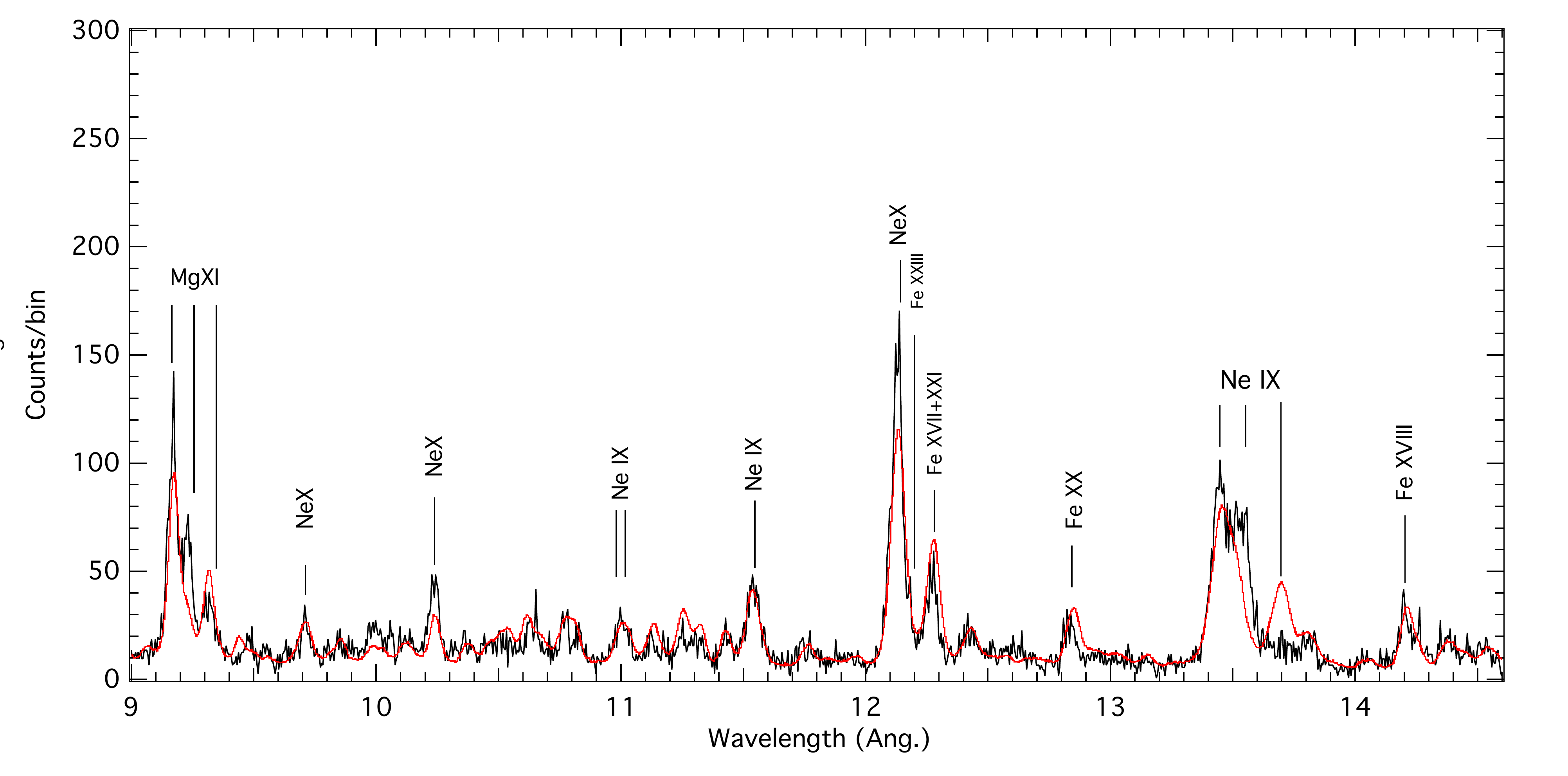}
   \includegraphics[width=5in]{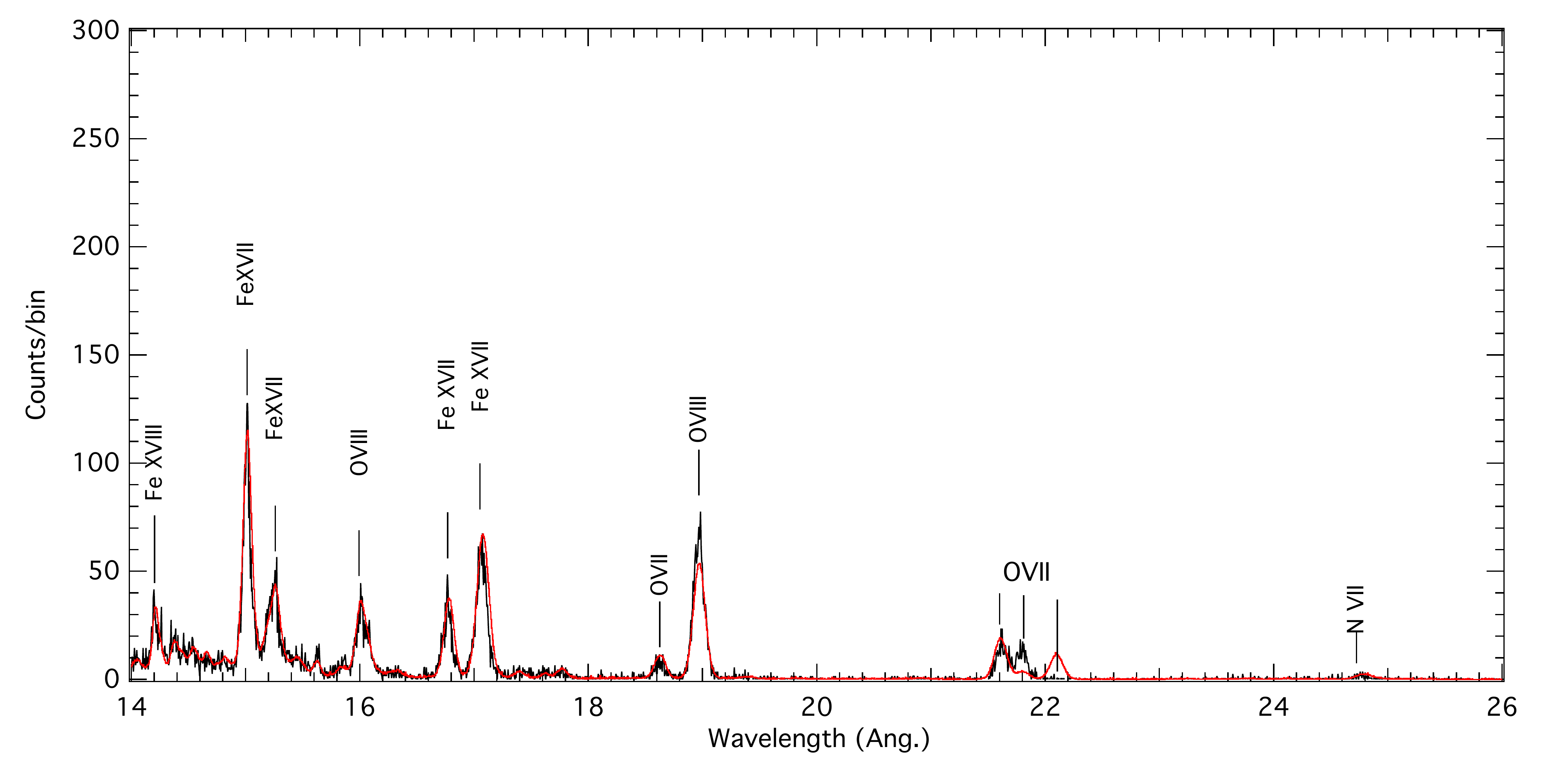}
   \caption{The combined  MEG+HEG spectrum of \dori\ (in black) with the 3-component fit (shown in red) given in Table \ref{tab:specfit}.  {The model spectra, which assume low density and do not include effects of UV photoexcitation,} generally overestimate the strength of the forbidden lines and underestimate the strengths of the intercombination lines, 
   especially at longer wavelengths, most notably at OVII.}
   \label{fig:specfit}
\end{figure*}

The derived temperature distribution is similar to that found by \cite{2002ApJ...577..951M} in their study of the 2000 January HETGS spectrum, and by \cite{Raassen:2013fk} in their analysis of an LETGS spectrum from 2007 November.  In general, aside from the overall weakness of the forbidden lines compared to the  model spectrum  {(which assumes a low-density plasma with no UV photoexcitation)}, the overall distribution of emission line strengths, and the continuum, are described reasonably well by the model.   {We note, in reality, that this three-temperature model is a simplified representation of the actual emission measure distribution with temperature. This multitemperature model mainly provides us with an adequate approximation of the local (pseudo-) continuum in order to improve line fitting and modeling.}

\subsection{Emission Lines}

The observed X-ray emission lines in our \dori\ spectrum provide important diagnostic information about the phase-averaged state of the hot gas within the wind of the system, and, as we show below, this is dominated by the shocked gas embedded within the wind of \dori a1, with little contribution (if any) from gas heated by the shock produced by the collision of the wind from \dori a1 with the wind or photosphere of \dori a2.  The analysis of the set of emission lines depends on choice of line profile, continuum level, and accounting for line blends.

\subsubsection{Gaussian Modeling}

To better account for blends and uncertainties in the continuum level, we performed a  Gaussian fit to the strong lines, allowing flux, line width, and centroid velocity to vary. These fits, shown in Figure \ref{fig:h-like}, were done using the three-temperature fit given in Section \ref{sec:temps} above to define the continuum and amount of line blending. We set the abundance of the element to be measured to zero, with the abundances of other elements set to solar and other parameters (temperature, absorptions) fixed at the values given in Section \ref{tab:specfit}. This procedure is useful to account for line blends, in particular, for the \ion{Ne}{10} line at 12.132~\AA, which is blended with an \ion{Fe}{17} line at 12.124~\AA. We assumed simple Gaussian line profiles for the line to be fit, and fit for both the Ly$\alpha1$ and Ly$\alpha2$ lines, with line widths and velocities fixed for both components, and the intensity ratio of the Ly$\alpha2$ to the Ly$\alpha1$ line set to the emissivity ratio at the temperature of peak emissivity. We used the Cash statistic and ISIS to perform the fits, simultaneously fitting the HEG and MEG $\pm1$ order spectrum from all four observations simultaneously.  Table \ref{tab:hlines} shows the result of fits of the H-like Ly$\alpha$ lines, plus the strong \ion{Fe}{17} line at 15.014~\AA. In general, the Gaussian fits are poor (the reduced Cash statistic $> 1.5$) except for the weak \ion{Si}{14} line,
though the asymmetries in the bright lines are not very strong. All of the line centroids are near zero velocity, though the \ion{Ne}{10} line is blue-shifted at about the 2--$\sigma$ level.

{\setlength{\extrarowheight}{3pt}
\begin{table*}[htp]
\caption{Gaussian Fits to the H-like lines, plus \ion{Fe}{17}}
\begin{center}
\begin{tabular}{lccrr}
\hline
    & \multicolumn{1}{c}{$\lambda$} & \multicolumn{1}{l}{Flux} & \multicolumn{1}{c}{V} & \multicolumn{1}{c}{HWHM} \\
    & \multicolumn{1}{c}{\AA} & \multicolumn{1}{c}{10$^{-5}$~ph. s$^{-1}$ cm$^{-2}$} & \multicolumn{1}{c}{\kms} & \multicolumn{1}{c}{\kms} \\ 
\hline
\ion{O}{8}   & 18.967
&$219_{-10}^{+9}$ & $-9_{-33}^{+37}$ & $918_{-29}^{+38} $ \\
\ion{Fe}{17} & 15.014
& $76_{-3}^{+4}$ & $-24_{-35}^{+42}$ & $971_{-27}^{+53} $ \\
\ion{Ne}{10}   & 12.132
&$10_{-1}^{+1}$ & $-102_{-42}^{+50}$ & $726_{-58}^{+48} $ \\
\ion{Mg}{12}   & 8.419
&$1_{-0}^{+0}$ & $-12_{-55}^{+33}$ & $547_{-61}^{+58} $ \\
\ion{Si}{14}   & 6.180
&$0.35_{-0.05}^{+0.05}$ & $-49_{-134}^{+45}$ & $544_{-124}^{+116} $ \\

\hline
\end{tabular}
\end{center}
\label{tab:hlines}
\end{table*}
\begin{figure*}[htbp]    \centering
   \includegraphics[width=2in, angle=-90]{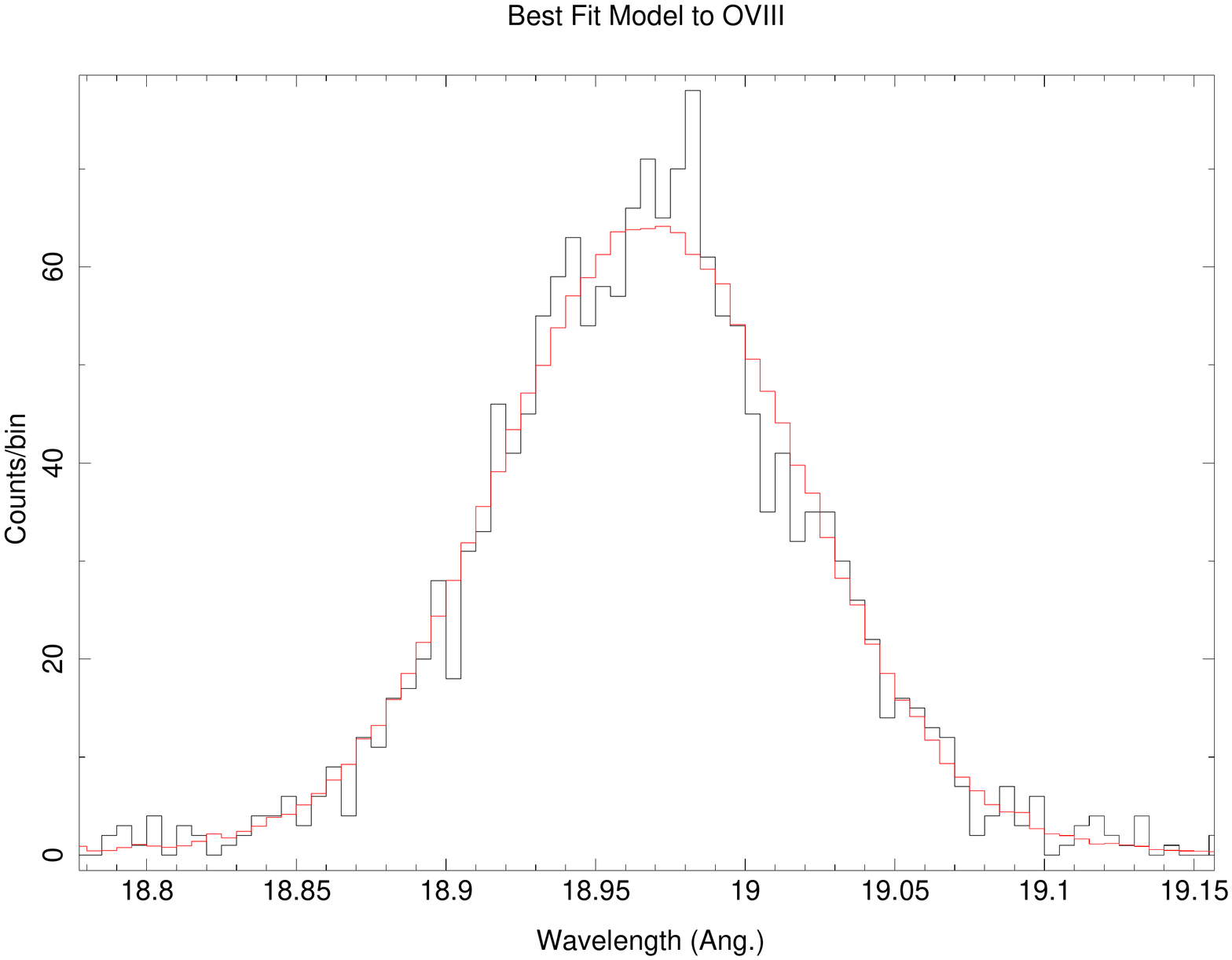}
   \includegraphics[width=2in, angle=-90]{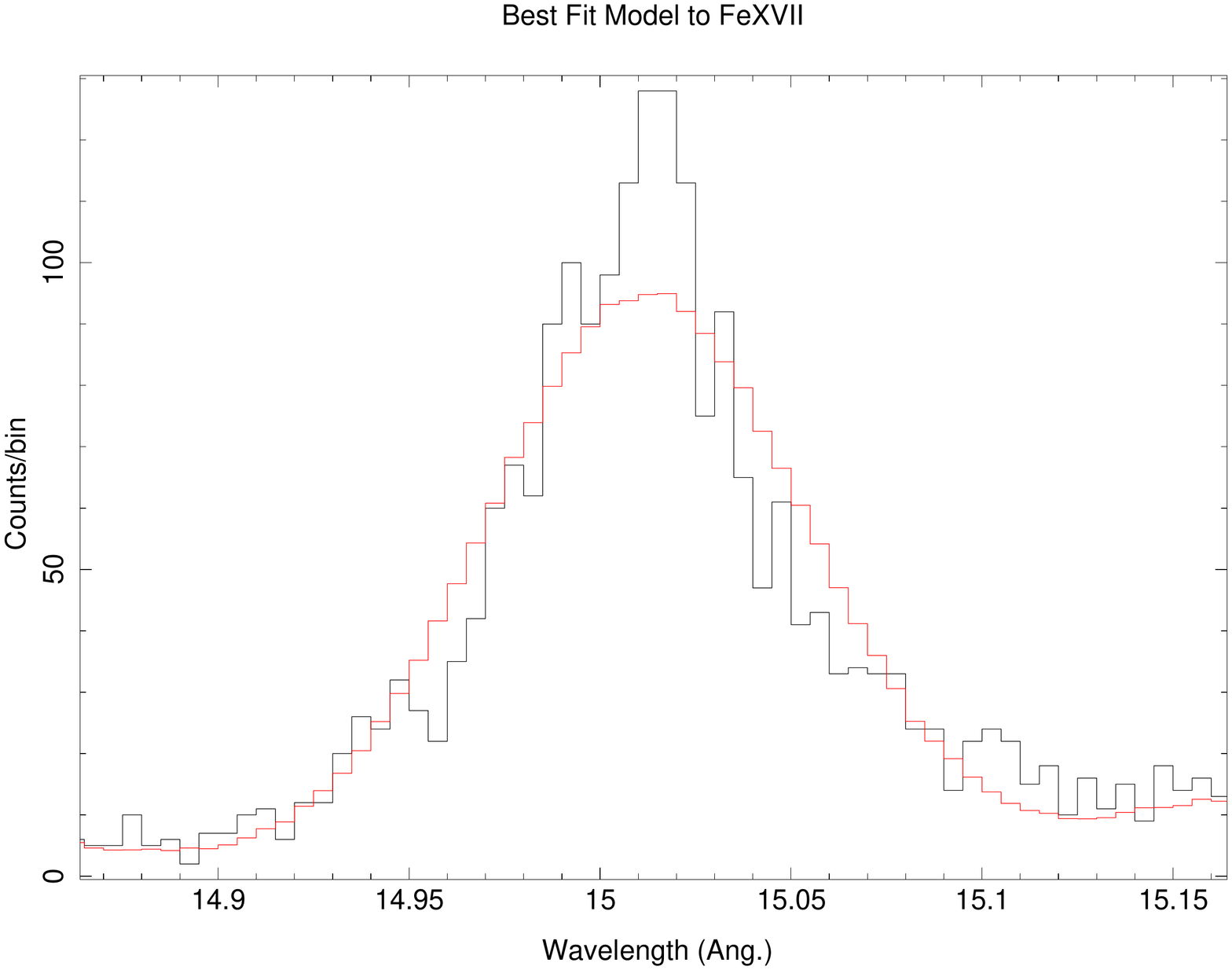}  
   \includegraphics[width=2in, angle=-90]{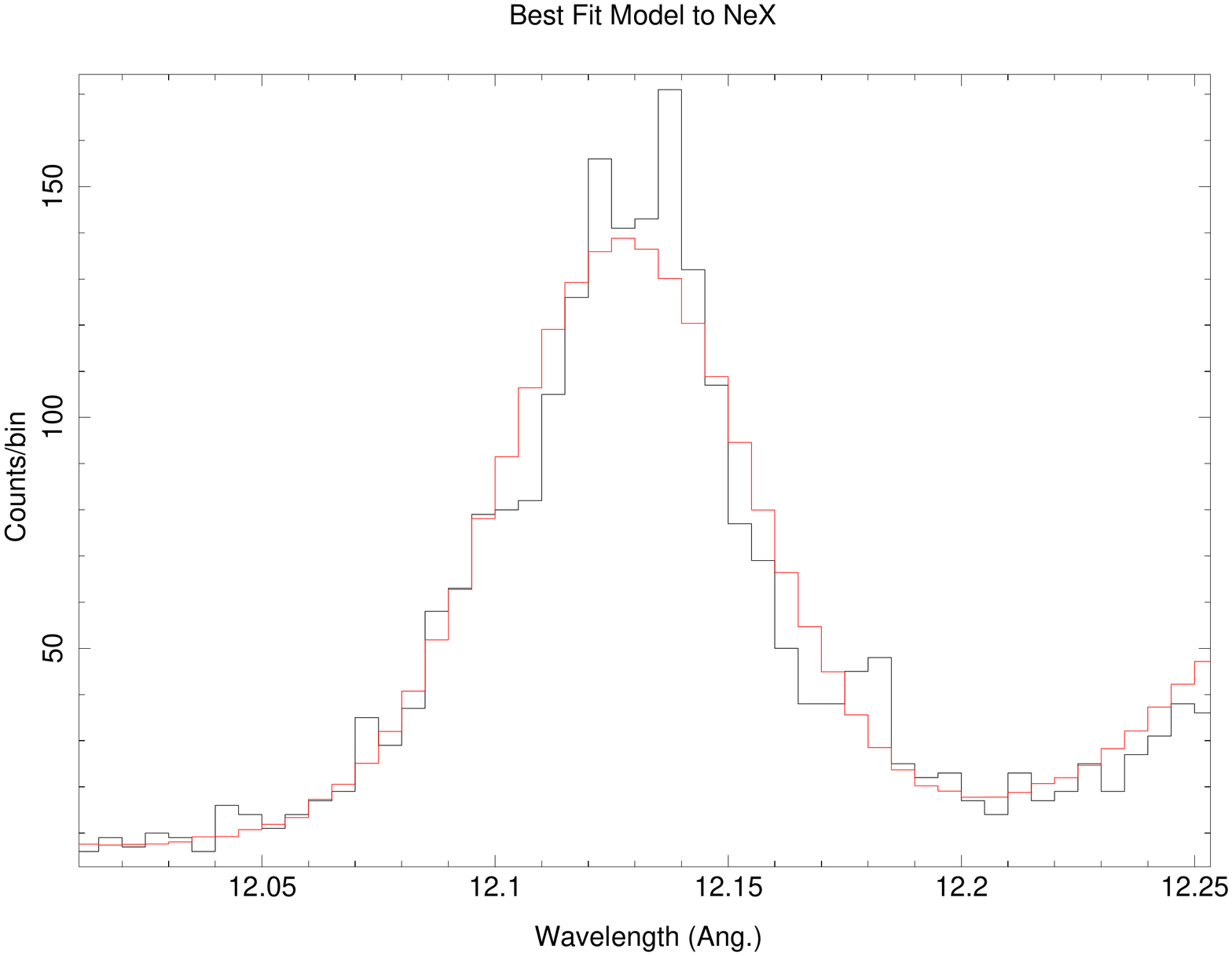} 
   \includegraphics[width=2in, angle=-90]{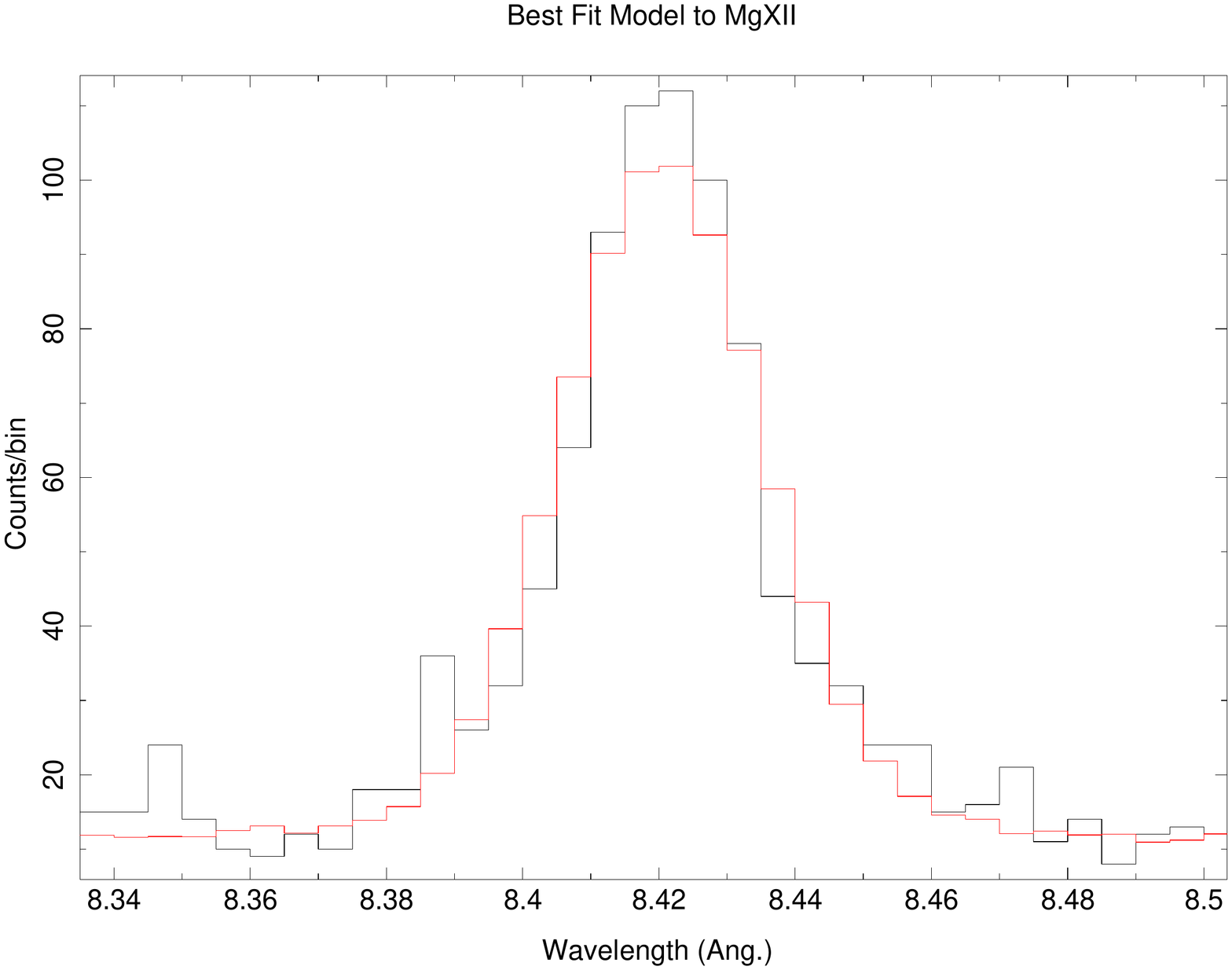} 
   \includegraphics[width=2in, angle=-90]{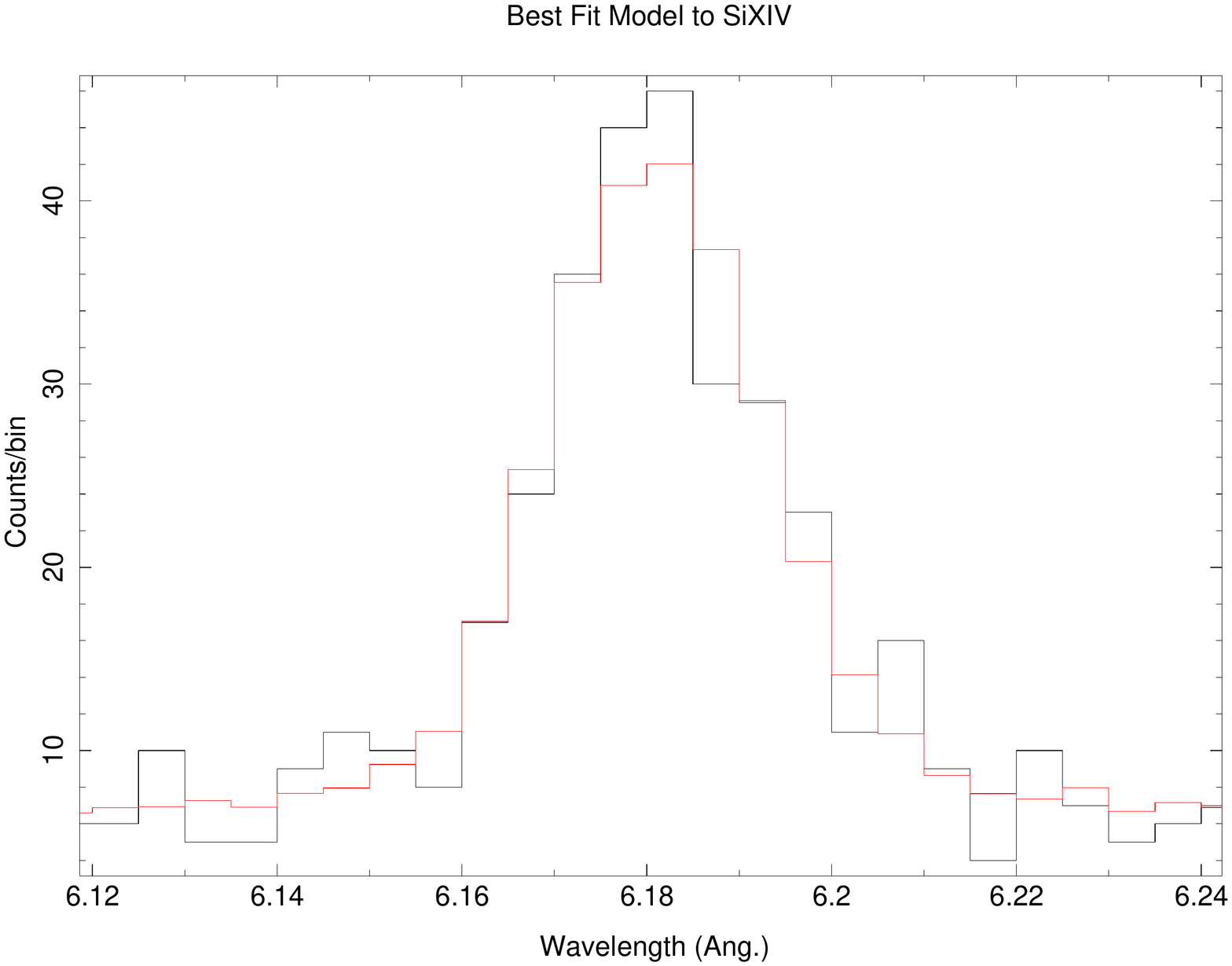} 
   \caption{Top to bottom, left to right: \ion{O}{8}; \ion{Fe}{17}; \ion{Ne}{10};  \ion{Mg}{12}; and \ion{Si}{14}.  The lines are plotted in the velocity range of $-3000$~\kms\ to $+3000$~\kms. The best-fit Gaussian profile, and the continuum derived from the model parameters given in Table \ref{tab:specfit} is shown in red. Note that while most of the Ly$\alpha$ lines are adequately described by a symmetric Gaussian, the \ion{Fe}{17} and \ion{Ne}{10} lines are not as well fit by simple Gaussian profiles as the other lines. This may be due to the effects of non-uniform X-ray line opacity, as discussed in Section \ref{sec:lineopacity}.}
   \label{fig:h-like}
\end{figure*}

\newpage

We also measured the forbidden ($z$), intercombination ($x+y$), and resonance components ($w$) above continuum for each of the helium-like ions (\ion{O}{7}, \ion{Mg}{10}, \ion{Ne}{9}, and \ion{Si}{13}) by Gaussian fitting. As before, we used the three-temperature fit given in Section \ref{sec:temps} above to define the local continuum near the line region.  Although the individual intercombination components ($x+y$) are unresolved in the HETGS spectra for all of the He-like ions, we included a Gaussian line for the $x$ and $y$ lines, but restricted the centroid velocity and line widths to be the same for both the $x$ and $y$ components. Because the forbidden, intercombination and resonance lines can have different spatial distributions throughout the wind, we allowed the widths, centroids, and line fluxes of these lines to vary individually.  The forbidden component of the \ion{O}{7} line is weak, and, in addition, this line was only observed in the MEG-1 spectrum arm because ACIS-S chip S5 was turned off due to spacecraft power constraints. To increase signal to noise for the \ion{O}{7} forbidden line, and for the weak \ion{Si}{13} and \ion{S}{15} triplets, we included data from the 2001 HETG and 2008 LETG observations when fitting.  Figure \ref{fig:he-gaussians} shows the fits to the He-like lines, and Table \ref{tab:helike} shows the results of this three-Gaussian component fitting, while Table \ref{tab:r-and-g} shows the $R=z/(x+y)$ and $G=(x+y+z)/w$ ratios.

\newpage
\begin{table*}[htp]
\caption{Gaussian fits to the He-like lines}
\begin{center}
\begin{tabular}{cccllll}
\hline
& \multicolumn{3}{c}{Centroid Velocity (\kms)} & \multicolumn{3}{c}{HWHM (\kms)} \\
ion &  \multicolumn{1}{c}{$w$} & \multicolumn{1}{c}{$x+y$} & \multicolumn{1}{c}{$z$} & \multicolumn{1}{c}{$w$} & \multicolumn{1}{c}{$x+y$} & \multicolumn{1}{c}{$z$} \\
\hline
OVII    & $166 \pm 19$ & $-194 \pm 18$ & $-810 \pm 384 $ & $761
\pm 14$ & $826
\pm 40$ & $160
\pm 270$ \\ Ne IX   & $-146 \pm 166$ & $-410 \pm 231 $ & $ 441   \pm  466 $&$ 849
\pm 138
$&$ 1057
\pm 222
$&$ 1289
\pm 49
$ \\
Mg XI   & $8    \pm 74 $ & $31   \pm 109 $ & $ -63   \pm  270 $&$ 782
\pm 97
$&$ 584
\pm 146
$&$ 1302
\pm 386
$ \\
Si XIII & $42 \pm 64$	 & $88\pm 191 $ & $ -60 \pm 21$ & $488\pm 69$ &$ 704
\pm 361$ & $506
\pm 79$ \\ S XV  
   & $99 \pm 357$ & $1168 \pm 1203$ & $-27 \pm 633$  & $540
\pm  206$ & $966
\pm 1256$ & $69
\pm 1254$ \\ \hline
\end{tabular}
\end{center}

\label{tab:helike}
\end{table*}

\newpage
\begin{table}[htp]
\caption{$R$ and $G$ ratios}
\begin{center}
\begin{tabular}{lcccc}
\hline
ion & $R=z/(x+y)$ & $G=(x+y+z)/w$  \\
\hline
O VII   & $0.04\pm0.01$  & $0.94\pm0.26$   \\
Ne IX   & $0.27 \pm  0.10$  & $1.44 \pm 0.65$   \\
Mg XI   & $0.96 \pm  0.36$  & $0.95 \pm 0.37$   \\
Si XIII & $1.77 \pm 0.18$  & $0.90\pm0.12$   \\ S XV    & $3.88\pm 2.86$  & $0.72\pm0.74$  \\

\hline
\end{tabular}
\end{center}
\label{tab:r-and-g}
\end{table}
\begin{figure*}[htbp]    \centering
   \includegraphics[width=3in]{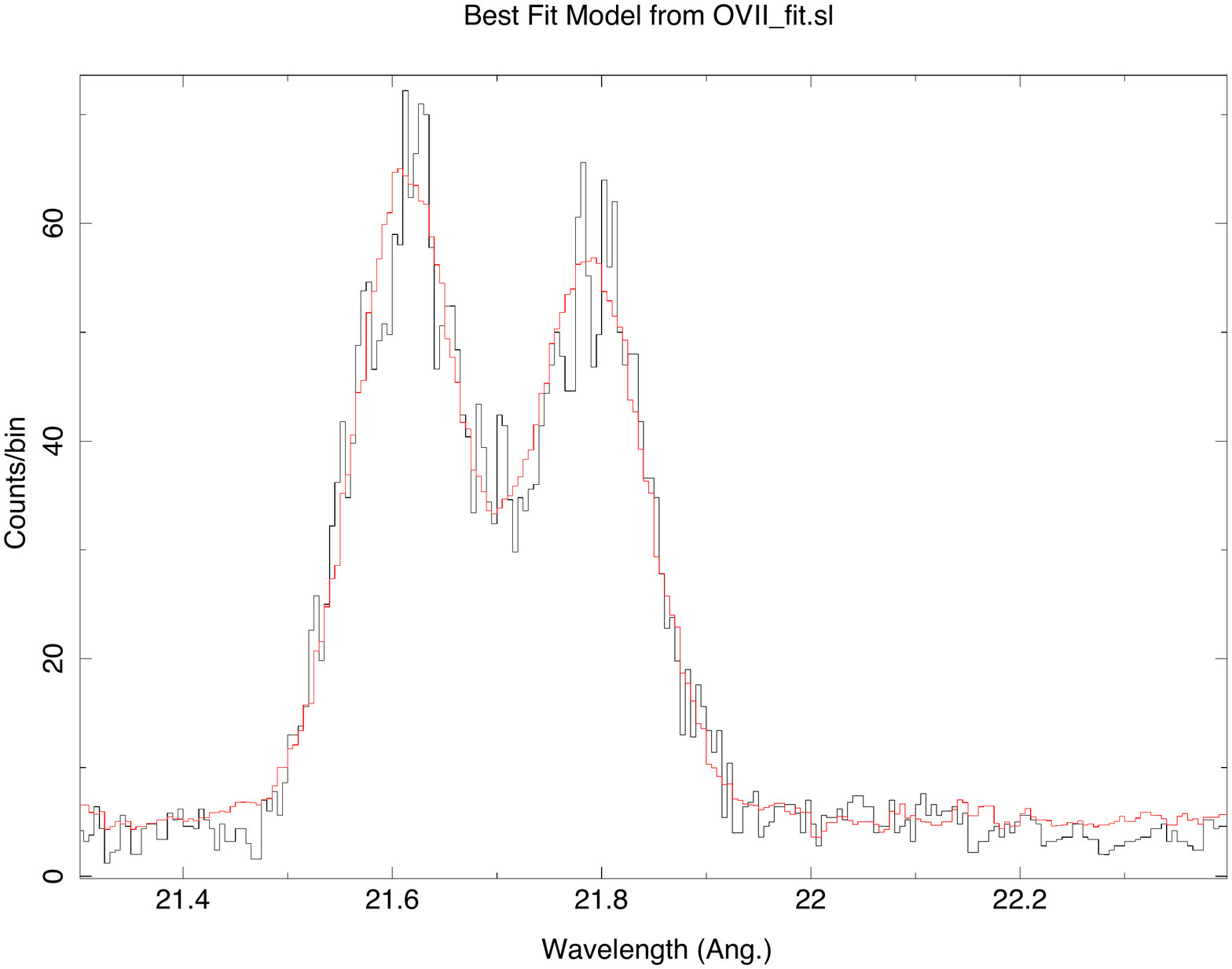} 
   \includegraphics[width=3in]{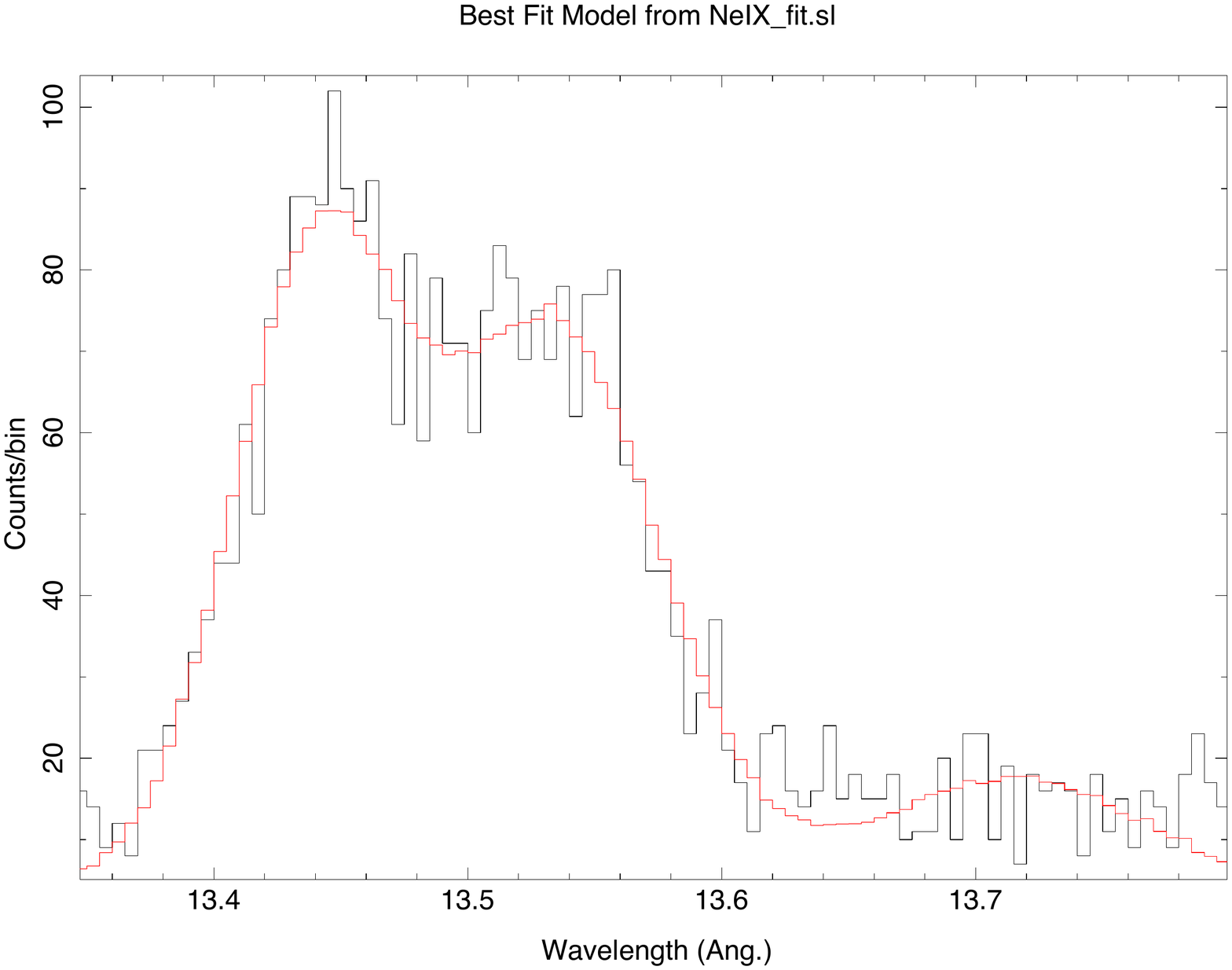} 
   \includegraphics[width=3in]{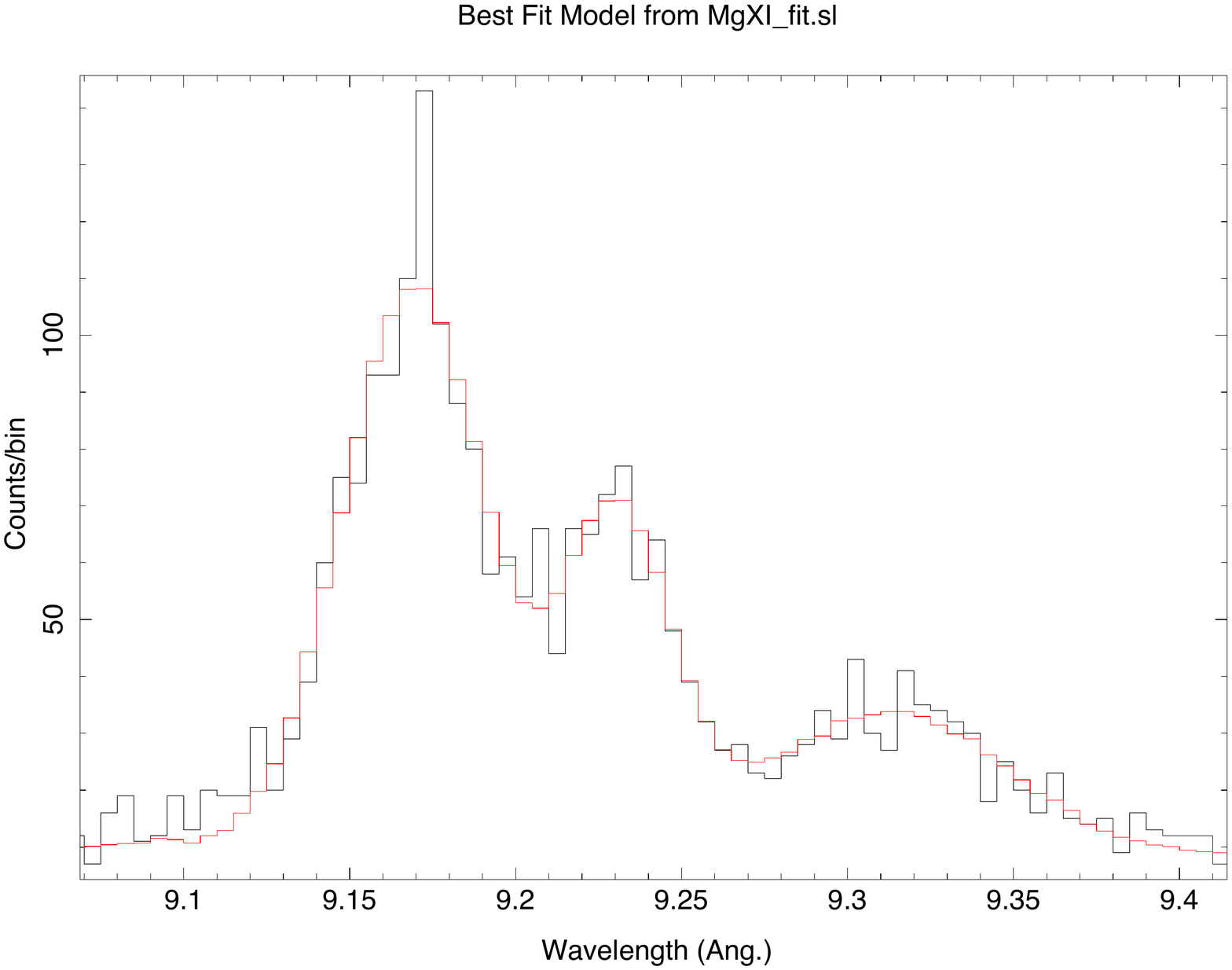} 
   \includegraphics[width=3in]{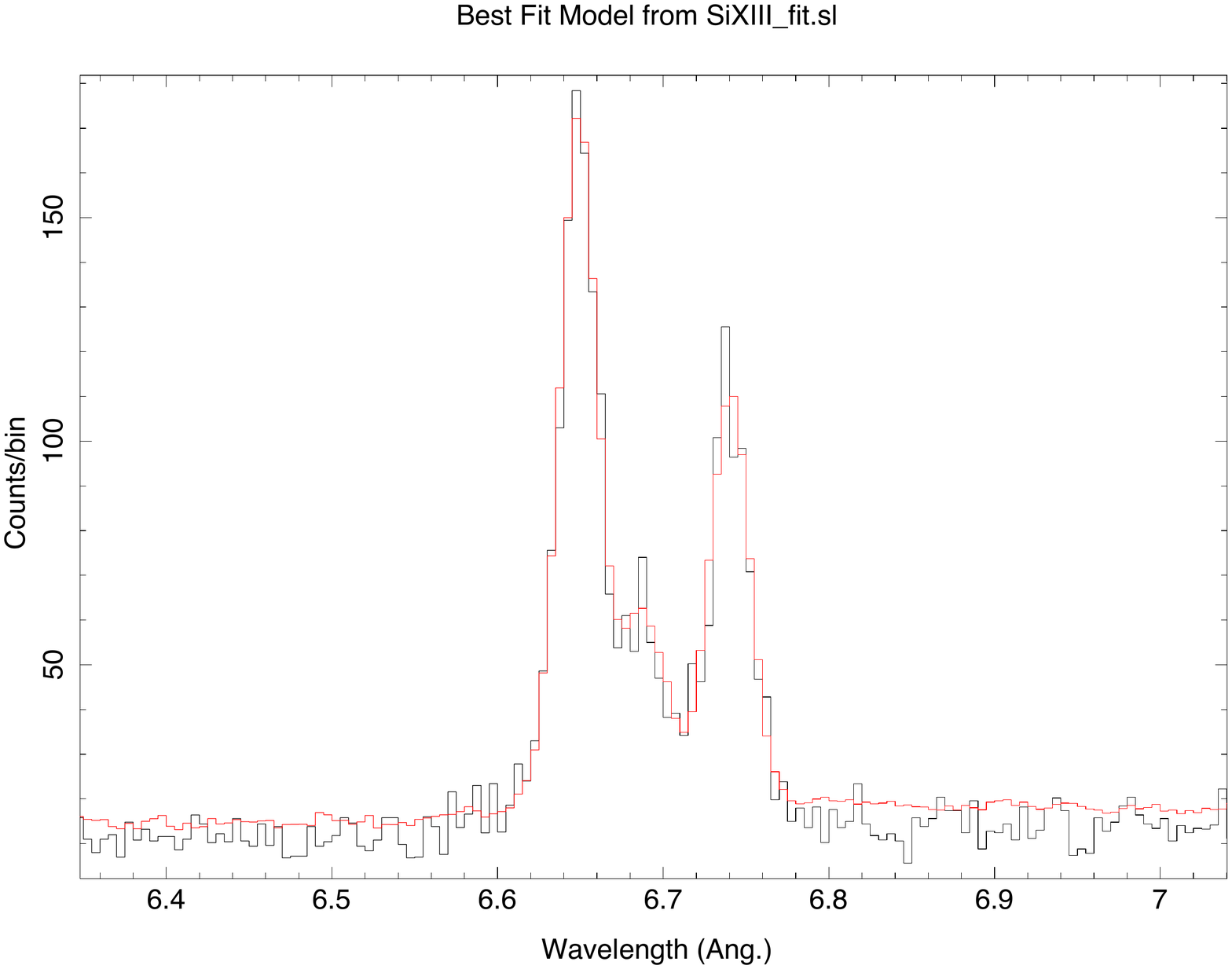}
   \includegraphics[width=3in]{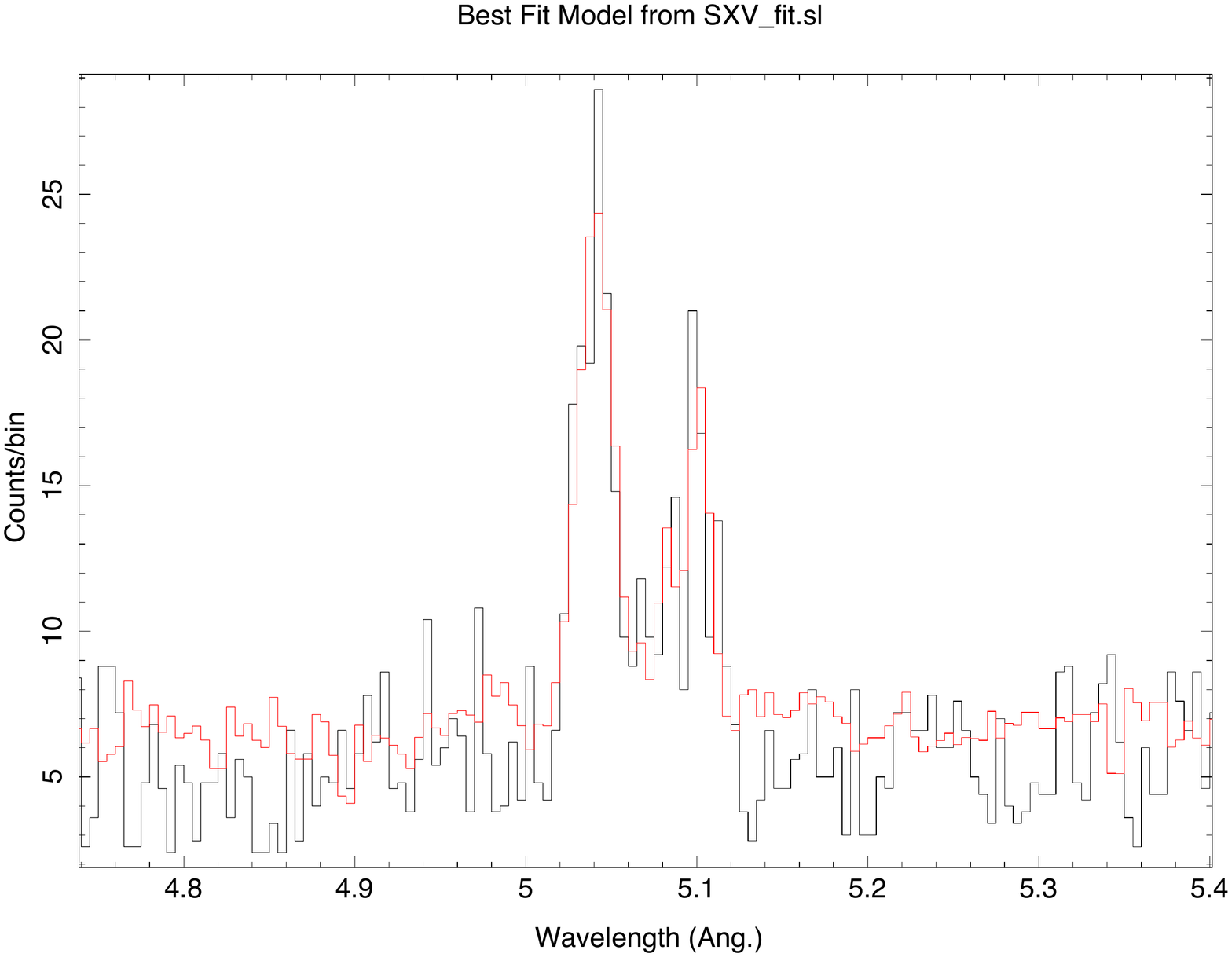}
   \caption{Top to bottom, left to right: O VII; Ne IX; Mg XI; Si XIII; S XV.  The best fit, using a model of 4 Gaussian lines ($w, x, y, \&~ z$ components) and the continuum derived from the model parameters given in Table \ref{tab:specfit}, is shown in red.}
   \label{fig:he-gaussians}
\end{figure*}

\begin{figure*}[htbp]    \centering
   \includegraphics[width=6in]{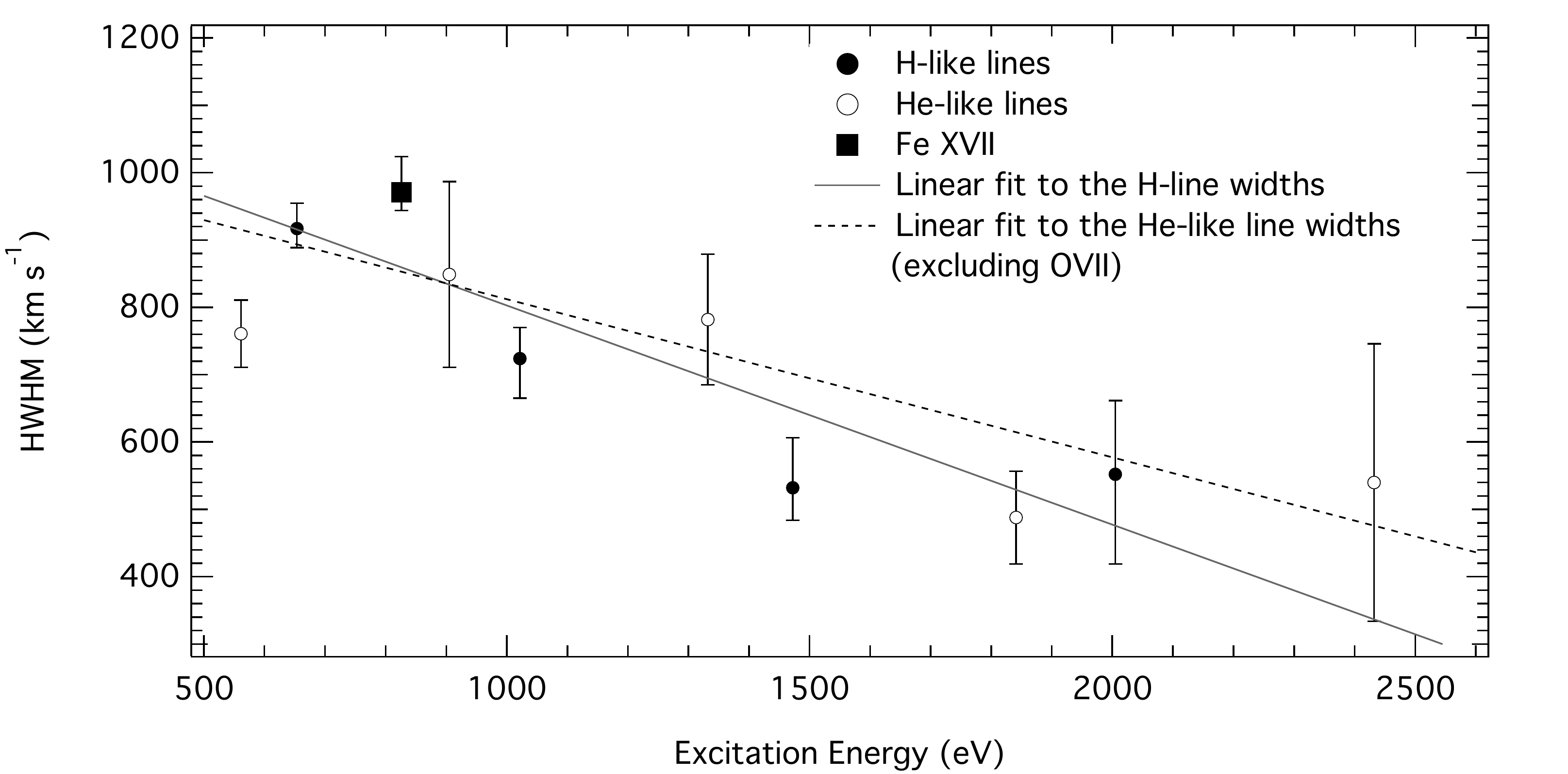} 
   \caption{Half-widths of the H-like Ly$\alpha$ lines (\kms) and the He-like resonance lines versus excitation energy (eV) of the upper level of the transition. The full and dashed lines represent the best linear fit to the HWHMs from the H-like lines, and the He-like lines (excluding the \ion{O}{7} width), respectively.}
   \label{fig:h-sigmas}
\end{figure*}
\newpage

Figure \ref{fig:h-sigmas} shows the dependence of the half width at half maximum of the Gaussian fit versus the excitation energy of the upper level of the transition.  
 {The linear correlation coefficient for the H-like half-widths is $-0.89$, indicating a strong anti-correlation between line half-width and excitation energy. For the He-like lines, the linear correlation coefficient is $-0.81$, also indicating a strong anti-correlation of line half-widths and excitation energy.}  
 {Thus the}
line widths are  {anti-}correlated with the upper energy level, in that the line width decreases with excitation energy.   {This anti-correlation shows that the more highly excited lines form at lower velocities, and thus closer to the stellar surface of the primary, indicating that the higher-temperature X-ray emission emerges from deeper regions in the wind than the cooler emission.}

In Figure \ref{fig:h-sigmas}, the \ion{O}{7} line width seems lower compared to the trend defined by the more highly excited ions. 
Excluding the \ion{O}{7} line, a linear fit to the remaining He-like lines yields a linear correlation coefficient of $-0.87$, indicating a stronger anti-correlation, and also results in a steeper linear slope.  This linear fit predicts that the \ion{O}{7} line should have a half-width of 918
eV, a factor of 1.2
larger than observed. We caution that, unlike the other lines, the \ion{O}{7} line was only observed in one grating order since ACIS-S chip 5 was switched off during these observations.

As a crude approximation, if we assume that the X-ray emitting material resides in a thin spherical shell at radius $r$ around \dori a1, then the line profile  will extend from $-V(r)$ to $+V(r)\sqrt{1-(R_{Aa1}/r)^2}$, where $R_{Aa1}$ is the radius of \dori a1, and $V(r) = V_{\infty, Aa1}(1-R_{Aa1}/r)^{\beta}$, the standard velocity law for radiatively driven winds.  The inverse correlation of the line widths with excitation energy suggests that the hotter X-ray emitting gas is formed over a smaller volume in the wind acceleration zone closer to the star, where wind 
radial velocity differentials are larger and where higher temperature shocks can be generated; cooler ions can be maintained 
farther out in the wind where the acceleration (and thus the velocity differential) is smaller. A similar conclusion was reached by \cite{Herve:2013rt} in their analysis of $\zeta$ Puppis.

\subsubsection{Effects of X-ray Line Opacity}
\label{sec:lineopacity}
 {The possibility that strong resonance line photons might be scattered out of the line of sight has significant implications on our physical understanding of the X-ray emission from hot stars, especially in the interpretation of mass-loss rates derived from X-ray line profiles and abundances derived from X-ray line ratios.} 
Resonance scattering may be important for lines with high oscillator strengths and could, in principle, change the line shape or intensity ratios, though recent analysis by \cite{Bernitt:2012kx} suggested that our poor knowledge of the underlying atomic physics may play the dominant role in accounting for discrepancies in line intensities. \cite{2002ApJ...577..951M} focussed on the Fe XVII lines at 15.014~\AA\ and 15.261~\AA, which have oscillator strengths of 2.49 and 0.64, respectively. 
 Resonance scattering might significantly affect the 15.014~\AA\ emission line, which is one of the strongest lines in the \dori\ X-ray spectrum, while scattering should be unimportant for the weak 15.261~\AA\ line. \cite{2002ApJ...577..951M} found that the observed ratio of these two lines, as derived from their \chandra\ grating spectrum, was $I_{15.01}/I_{15.26}=2.4\pm1.3$, nominally (though not significantly) below the optically thin limit $I_{15.01}/I_{15.26}=3.5$ derived from the \citet{Smith:2000qv} version of the Astrophysical Plasma Emission Code (APEC). 

We re-examined this issue for these two \ion{Fe}{17} lines using our deeper spectrum and a slightly different technique. We isolated the Fe XVII line region in the combined spectrum and fit this restricted region with an APEC-derived model, with abundances fixed at solar, including line broadening.  
We first fit the \ion{Fe}{17} line at 15.261~\AA, ignoring the region around the stronger 15.014~\AA\ line. We then included the 15.014~\AA\ line region and compared the predicted strength of the model 15.014~\AA\ line to the observed line.  This technique, in which we use a full thermal model to fit the spectra rather than a simple comparison of line intensities, has the benefit that line blends in the region will be more properly taken into account.  We found that the model based on the best fit to the 15.261~\AA\ line greatly overpredicted the strength of the 15.014~\AA\ line, and can be ruled out at high confidence ($\chi^{2}_{\nu}=3.57$, restricted to the 14.90--15.14~\AA\ region; excluding this region, $\chi^{2}_{\nu}=0.72$).
This may be an indication of the effect of resonance scattering on the 15.014~\AA\ \ion{Fe}{17} line.  Since it appears that the 15.014~\AA\ line is a bit narrower than the 15.261~\AA\ line, we also re-did the fit, allowing the width of the 15.014~\AA\ line to differ from that of the 15.261~\AA\ line. We then re-fit only the 15.014~\AA\ line, allowing the line broadening to vary and also allowing the normalization to vary.  Figure \ref{fig:fexvii} shows the resulting fit.  The best-fit HWHMs for the 15.014~\AA\ and 15.261~\AA\ lines are 
$1275^{+48}_{-268}$~\kms\ and $1496^{+109}_{-113}$~\kms, respectively, while the model normalizations are 
$0.0024^{+0.0001}_{-0.001}$ and $0.0030^{+0.001}_{-0.001}$ for the 15.014~\AA\ and 15.261~\AA\ lines, respectively.  This analysis also shows the 15.014~\AA\ line is significantly weaker than expected compared to the 15.261~\AA\ line.  This again may indicate that resonance scattering plays a role in determining the line profile shape and line strength, at least for the \ion{Fe}{17} line, though uncertainties in the atomic models and in our definition of the temperature distribution for \dori\ may play a significant role in altering the intensity ratios for these lines.

\begin{figure*}[htbp]    \centering
   \includegraphics[height=3in, angle=-90]{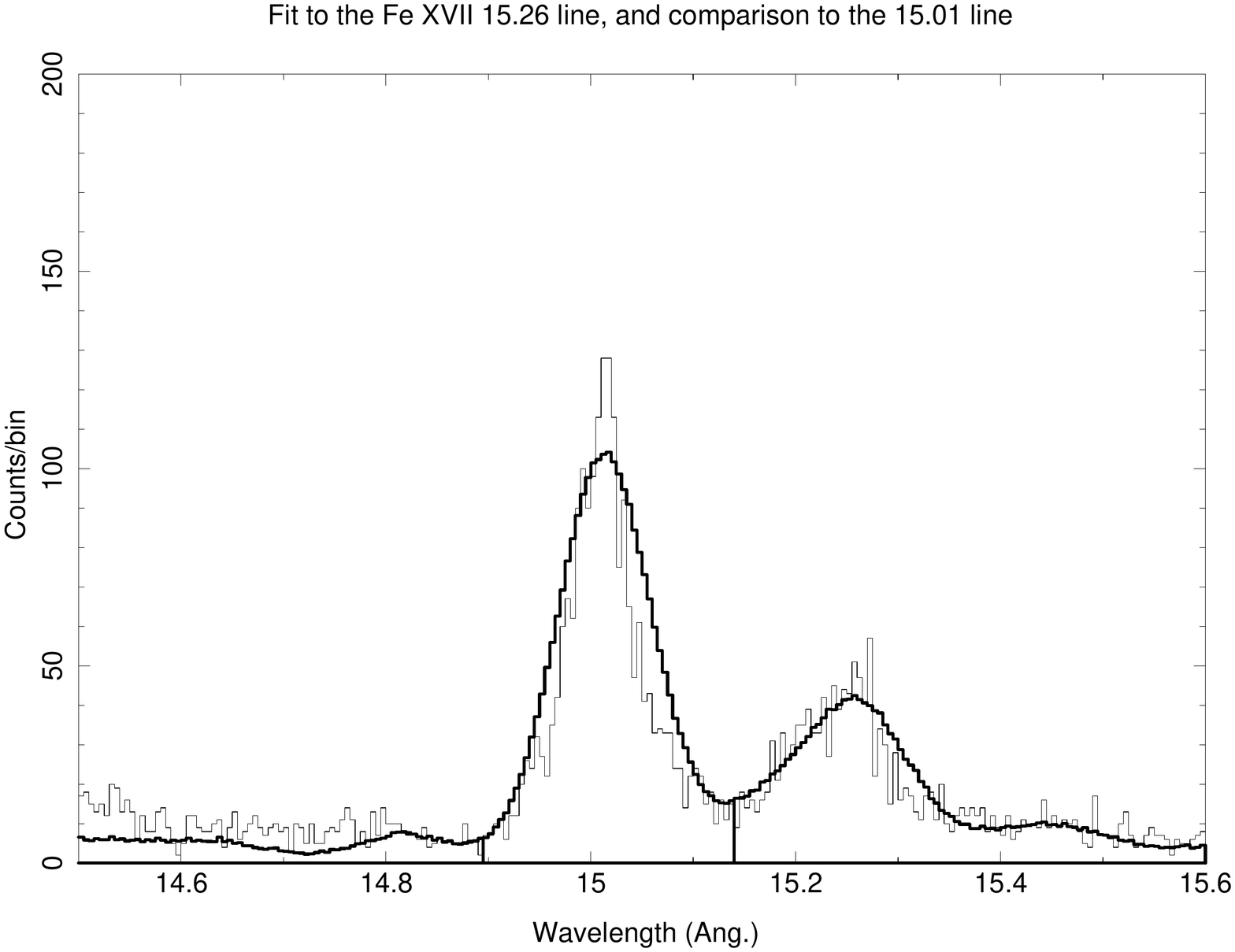} 
   \includegraphics[height=3in, angle=-90]{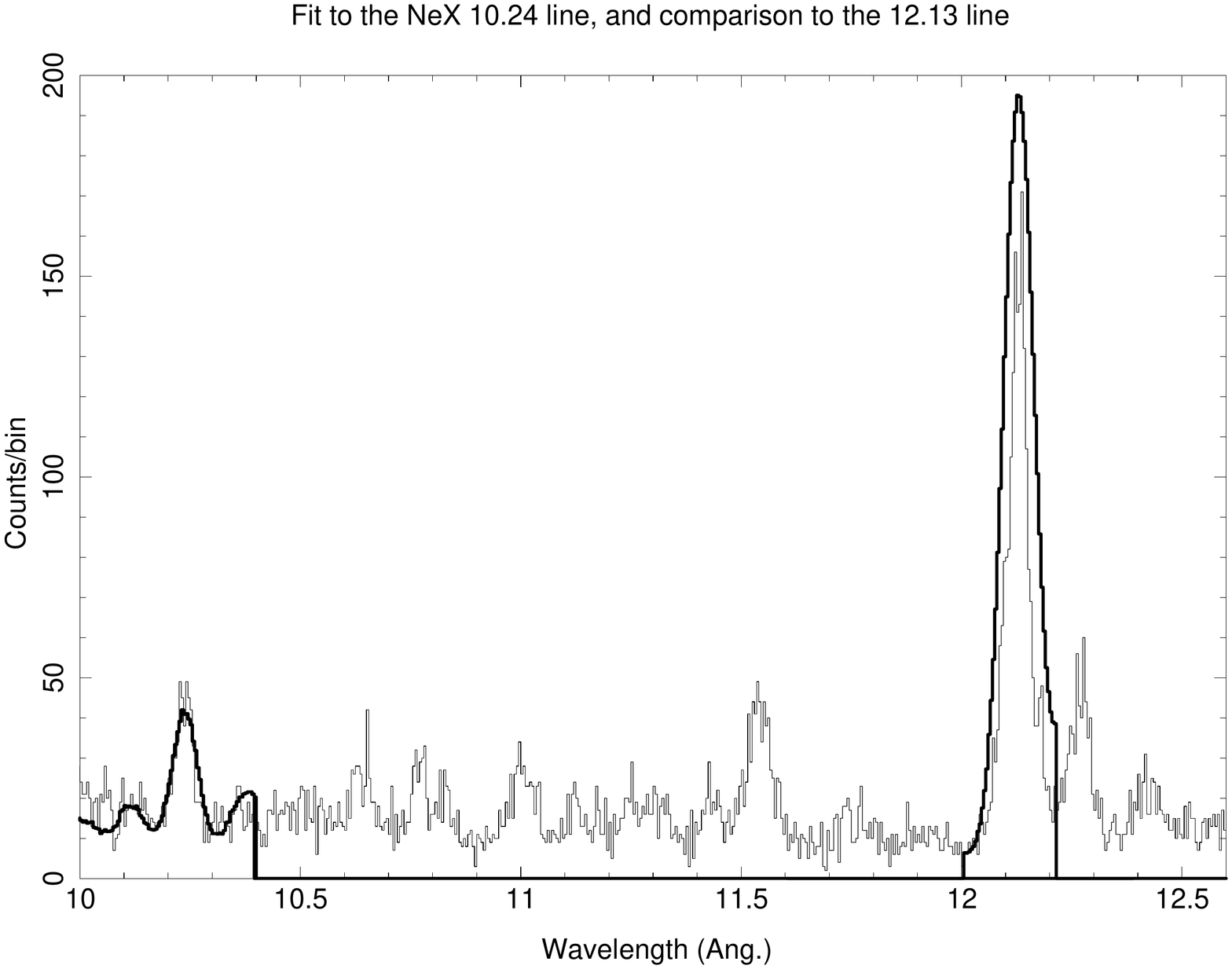}
   \caption{Left: APEC-based modeling of the \ion{Fe}{17} 15.014~\AA\ vs 15.261~\AA\ lines.  We first fit the 15.261~\AA\ line by itself. The thick histogram compares that model to the observed spectrum in the 14.5--15.6\AA\ range.  This shows that the model tgat fits the 15.26~\AA\ line overpredicts the strength of the 15.01\AA\ line. The vertical lines from the continuum to the X-axis at 14.9~\AA\ and 15.14~\AA\ show the adopted wavelength range of the 15.014~\AA\ FeXVII line. Right: APEC model fit to the \ion{Ne}{10} 10.24~\AA\ compared to the NeX 12.134~\AA\ line. The model (shown by the thick histogram) that fits the weaker 10.24~\AA\ line overpredicts the strength of the stronger 12.134~\AA\ line.}
   \label{fig:fexvii}
\end{figure*}

To further investigate the importance of resonance scattering, we also considered the \ion{Ne}{10} lines at 10.239~\AA\ and at 12.132~\AA, which have oscillator strengths of 0.052 and 0.28, respectively. These lines complement the \ion{Fe}{17} analysis since for \ion{Ne}{10} the stronger line appears at longer wavelength; this means that any effects of differential absorption that might affect the \ion{Fe}{17} line analysis would have the opposite effect on the \ion{Ne}{10} lines. We again fit the \ion{Ne}{10} 10.239~\AA\ line with a single temperature APEC model, but fixed the temperature to the temperature of maximum emissivity of the \ion{Ne}{10} lines, i.e. $T=6.3\times10^{6}$~K.  We then compared the model that best fits the \ion{Ne}{10} 10.239~\AA\ line to the \ion{Ne}{10} 12.132~\AA\ line. Note that the \ion{Ne}{10} 12.132~\AA\ line is blended with the \ion{Fe}{21} line at 12.285~\AA\ (which has a temperature of maximum emissivity of $12.6\times10^{6}$~K, about twice that of the \ion{Ne}{10} line), so we restricted the \ion{Ne}{10} 12.132~\AA\ fitting region to the interval 12.0--12.22~\AA.  We again find that the model, which provides a good fit to the weaker line ($\chi^{2}_{\nu}=0.79$), overpredicts the strength of the stronger line ($\chi^{2}_{\nu}=8.63$), again a possible indication that resonance scattering is important in determining the flux of the strong line.

\section{The Influence of Colliding Winds on the Embedded X-ray Emission}
\label{sec:cw}

\begin{figure*}
  \centering
  \includegraphics[width=3in]{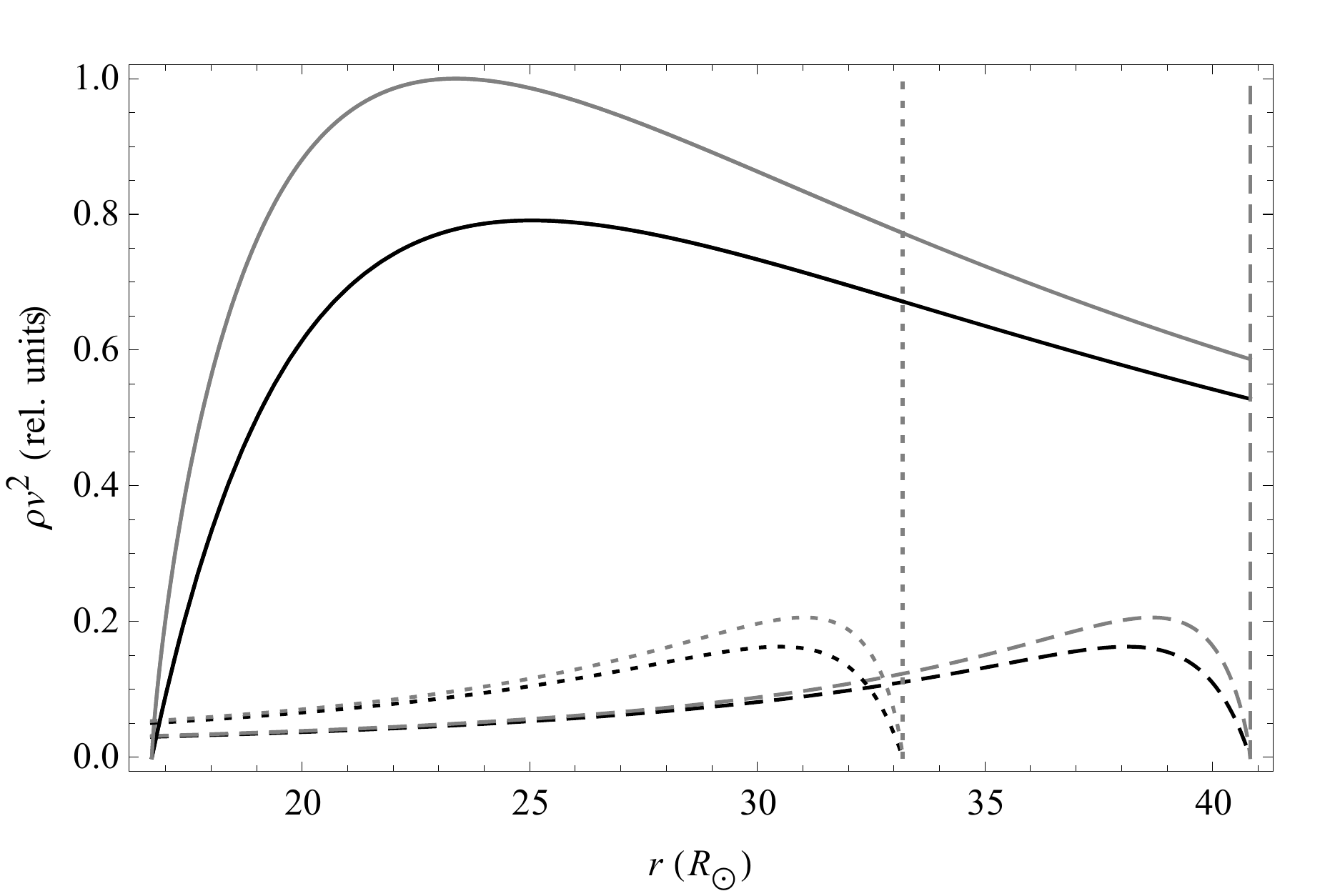}
  \includegraphics[width=3in]{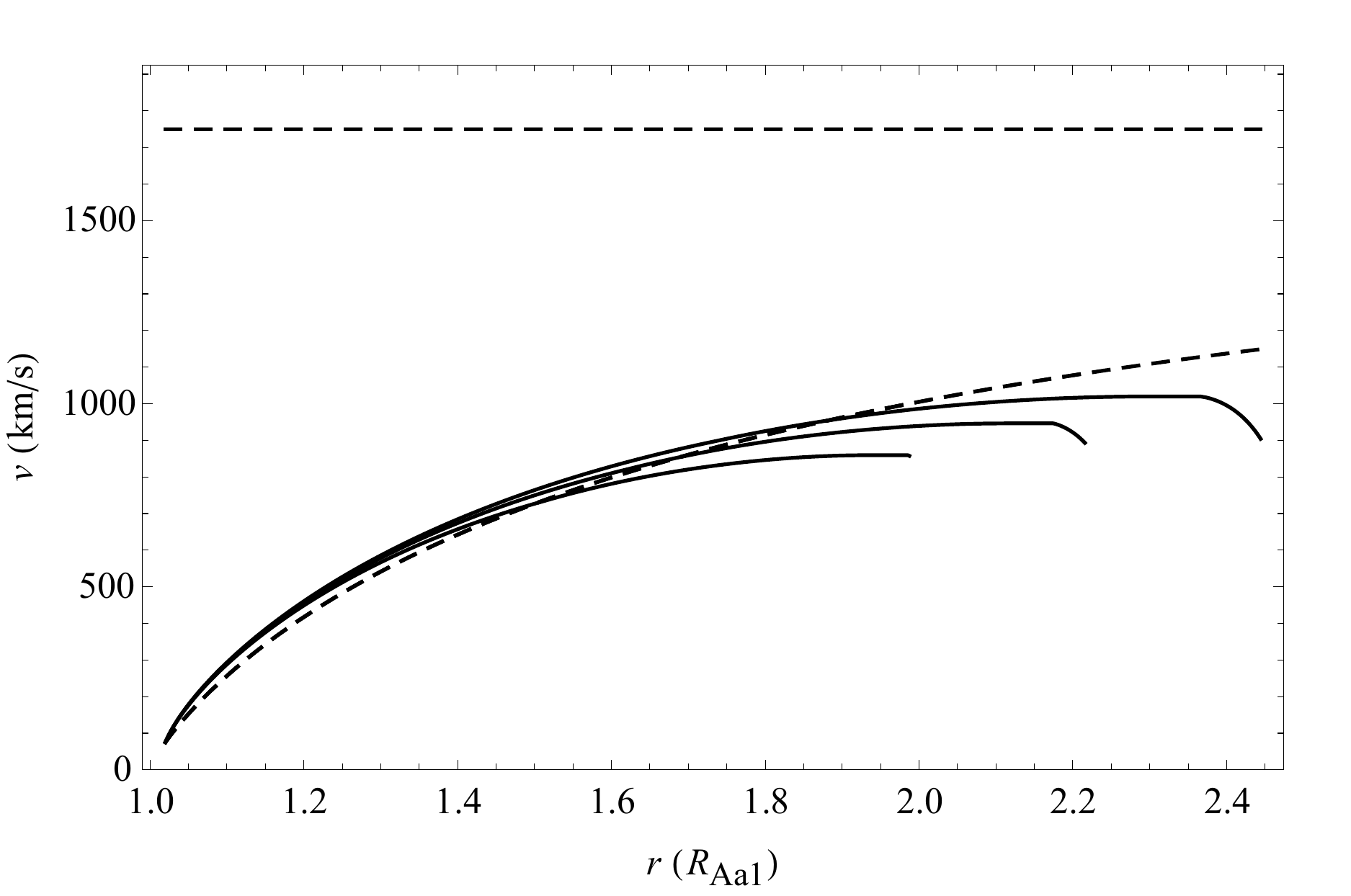}  
  \caption{Left: Ram pressure of Aa1 (solid) and Aa2 at apastron (dashed) and periastron (dotted).  The black lines show a $\beta$=1 law, while the gray lines show a $\beta$=0.8 law.  The gray vertical lines represent the location of Aa2's surface for these two phases. Right: 1D solution to the equation of motion of the primary wind along the line between the stars (solid) at three different separations --- apastron (top), semi-major axis (middle), and periastron (bottom).  For comparison, the dashed curve shows a $\beta$=0.8 law, and the dashed line shows terminal velocity.}
  \label{fi:RPB}
\end{figure*}

\begin{figure*}
  \centering
  \includegraphics[height=65mm]{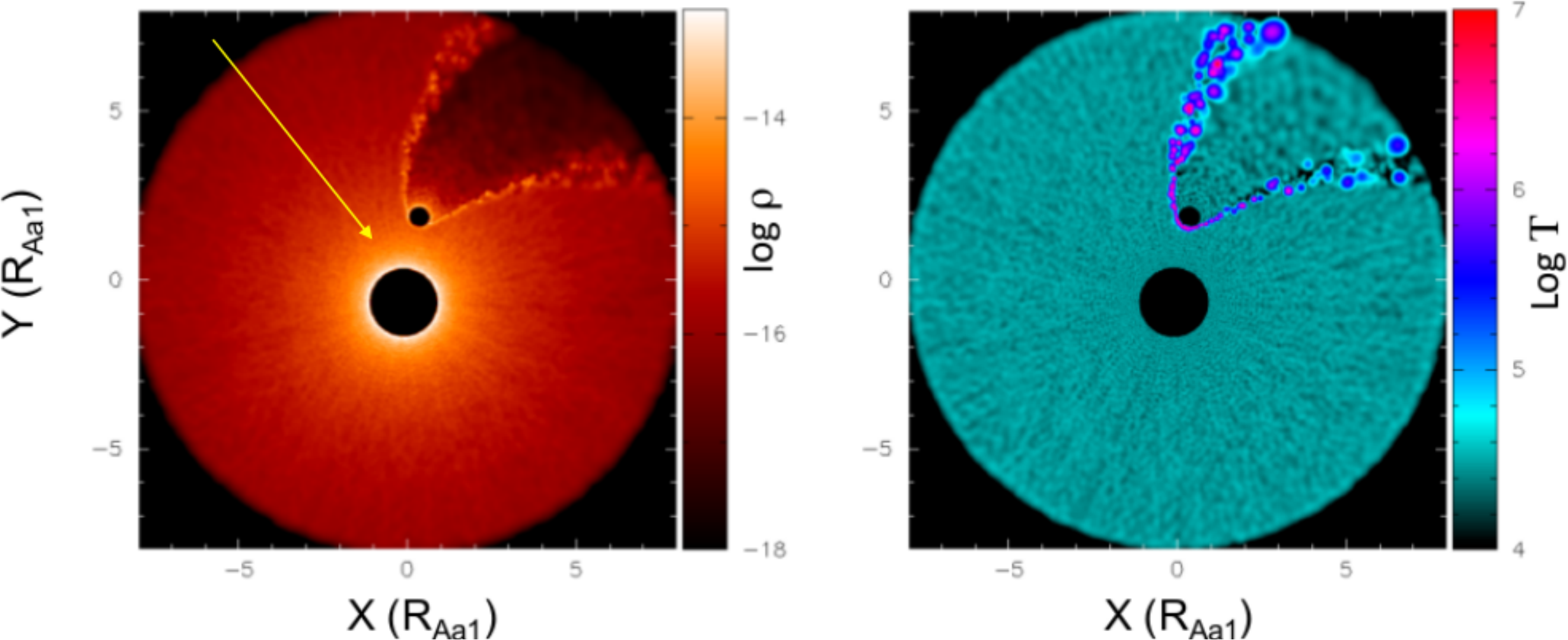}  \caption{Density (left) and temperature (right) structure in the orbital plane from the SPH simulation of \dori a1 (larger star) and \dori a2 (smaller star).   {The arrow shows the orientation of the line of sight.  The system is pictured at  phase $\phi=0.87$}. The collision of the wind from \dori a1 against \dori a2 produces a low-density cavity in the wind of \dori a1, where the emission from embedded wind shocks is reduced.  The collision also produces a layer of hot shocked gas at the boundary of the cavity which produces $<10$\% of the emission from the wind shocks embedded in the unperturbed wind from \dori a1. In the temperature plot on the right, the hot gas from embedded wind shocks in the winds from \dori a1 and \dori a2 is ignored, to emphasize the hot gas along the wind collision boundary.  }
  \label{fi:DT1}
\end{figure*}

Colliding winds can have important observable effects in our analysis of the X-ray emission from $\delta$ Ori Aa in two ways. The collision of the primary wind with the surface or wind of the secondary could produce hot shocked gas which might contaminate the X-ray emission from the embedded wind shocks in the primary's unperturbed wind. In addition, the colliding wind ``bow shock'' around the weaker-wind secondary produces a low-density cavity in the primary wind, and this cavity, dominated by the weak wind of $\delta$ Ori Aa2, should show little emission from embedded wind shocks. 
Along the line between the stars, the stellar winds will collide at the point at which their ram pressures $\rho v_\perp^2$ are equal \citep[e.g.][]{StevensBlondinPollock92}.  Using the stellar, wind, and orbital parameters in Table \ref{tab:params}, Figure \ref{fi:RPB} shows the ram pressures for Aa1 (solid) and Aa2 (dashed: apastron, dotted: periastron) assuming that the wind from each star follows the standard $\beta$ velocity law, $V(r)=v_\infty(1-R/r)^\beta$, where $V(r)$ is the wind radial velocity at a distance $r$ from the star, $R$ is the stellar radius, and we assume that $\beta=0.8$ or $1.0$.
The ram pressure of Aa1's wind is greater than that of Aa2 throughout the orbit, so the wind from Aa1 should directly impact  Aa2's surface, in this simple analysis.

A more thorough treatment includes the effects of Aa2's radiation on the wind of Aa1 (and vice versa).  These effects include ``radiative inhibition'' \citep{StevensPollock94} in which Aa1's wind acceleration along the line between the stars is reduced by Aa2's radiative force acting in opposition to the wind flow, and ``sudden radiative braking'' \citep{OwockiGayley95,GayleyOwockiCranmer97}, where Aa1's strong wind, which would otherwise impact the surface of Aa2, is suddenly decelerated by Aa2's radiation just above the surface of Aa2.  To estimate the magnitude of these effects, we solve the 1D equation of motion along the line of centers, accounting for both star's radiative forces via the standard Castor, Abbott, and Klein (CAK) line forces \citep{CastorAbbottKlein75} including the finite disk correction factor \citep{FriendAbbott86,PauldrachPulsKudritzki86} and gravitational acceleration.
We determine the CAK parameters $\bar Q$ and $\alpha$ \citep{Gayley95} to yield the desired mass-loss rates and terminal speeds for each star by using the standard reduction in mass-loss rate from the finite disk correction factor, i.e., $\dot M_{\rm fd}=\dot M_{\rm CAK}/(1+\alpha)^{(1+\alpha)}$.  We numerically integrate the equation of motion to distances far from the star to yield the terminal velocity.  Then we repeat the process including the radiation and gravity of both stars to determine the speed of each wind along the line between the stars.

Figure \ref{fi:RPB} shows the equation-of-motion solution for the primary wind.  The initial velocity corresponds to a $\beta=0.8$ law, but radiative inhibition causes the wind (solid) to accelerate less compared to the unmodified $\beta$-law (dashed).  In addition, the primary wind velocity does begin to decrease from radiative braking.  However, Star Aa2's surface is located at the end of each line, so that the primary wind does not completely stop before it impacts the secondary surface.  This indicates that the wind from star Aa1 should still impact the surface of Aa2, even when the influence of the radiation field of star Aa2 is taken into account.  Furthermore, due to the strong radiation of Aa1, the wind of Aa2 does not accelerate off the surface of the star toward Aa1, further suggesting that Aa1's wind will directly impact Aa2's surface.

We used a 3D smoothed particle hydrodynamics (SPH) code developed by \cite{Benz90} and \cite{BateBonnellPrice95} to model the effects of the wind--wind collision on the extended system wind.  \citet{OkazakiP08} was the first to apply  this code to a colliding-wind system, and \citet{Russell13} and \citet{MaduraP13} describe the current capabilities of the code, which we briefly state here.
The stars are represented as two point masses, and throughout their orbit they inject SPH particles into the simulation volume to represent their stellar winds.  The SPH particles are accelerated away from their respective stars according to a $\beta$=1 law (absent from any influence from the companion's radiation) by invoking a radiative force with a radially varying opacity $\kappa(r)$, i.e.\ $g_{\rm rad}=\kappa(r)F/c$, where $F$ is the stellar flux.
We take effects of the occultation of one star's radiation by the other star into account.  Radiative inhibition is included in the code (within the context of the radially varying opacity method), but radiative braking is not since it requires the full CAK solution for the wind driving, which is not yet included in the SPH code.  Radiative cooling is implemented via the Exact Integration Scheme \citep{Townsend09}, and the abundances of both winds are assumed to be solar \citep{AsplundP09}.

The importance of radiative cooling of the shocked material is determined by the parameter $\chi=d_{12}v_8^4/\dot{M}_{-7}$ \citep{StevensBlondinPollock92}, where $d_{12}$ is the distance to the shock in 10$^{12}$\,cm, $v_8$ is the preshock velocity in 10$^8$\,cm\,s$^{-1}$, and $\dot{M}_{-7}$ is the mass-loss rate in 10$^{-7}$\,$M_\odot$\,yr$^{-1}$.  $\chi>1$ indicates adiabatic expansion is more important, while $\chi<1$ indicates that the shocked gas will cool radiatively.  For the $\beta$=1 law, $\chi$ ranges from $0.5\lesssim\chi\lesssim1.3$ between periastron to apastron, so the shocked gas should cool through a combination of adiabatic expansion and radiation. 

Figure \ref{fi:DT1} shows the density and temperature structure of the interacting winds in the orbital plane  using the parameters in Table \ref{tab:params}.  The primary wind impacts the secondary star as expected from the analytical treatment above, where it shocks with newly injected secondary SPH particles.  If this interaction leads to SPH particles, either belonging to Aa1 or Aa2, going within the boundary of the secondary star, these particles are accreted, i.e.\ removed from the simulation.  The temperature plot of figure \ref{fi:DT1} shows that this leads to hot, shocked gas around Aa2, but this must be deemed approximate since the code does not force the Aa1 particles to accrete at the sound speed, which would increase the shock temperature, nor does it include any reflection of Aa1's radiation off of the surface of Aa2, which would decrease the shock temperature.  The half-opening angle is $\sim30^\circ$, so $\sim8\%$ of the solid angle of Aa1's wind is evacuated by Aa2 and its wind.

To determine the X-ray flux from the wind--wind/wind--star collision, we solve the formal solution to radiative transfer along a grid of rays through the SPH simulation volume, for which we use the SPH visualization program \texttt{Splash} \citep{Price07} as our basis.  The emissivity is from the \texttt{APEC} model \citep{SmithP01} obtained from \texttt{XSPEC} \citep{Arnaud96}, the circumstellar material absorbs according to the \texttt{windtabs} model \citep{LeuteneggerP10}, and the interstellar absorption is from \texttt{TBabs} \citep{WilmsAllenMcCray00}. 
The radiative transfer calculation is performed at 170 energies logarithmically spaced from 0.2 to 10 keV (100 per dex), and generates surface brightness maps for each energy.  These are then summed to determine the model spectrum, and finally folded through X-ray telescope response functions to directly compare with observations.
The overall contamination level of wind--wind/wind--star collision X-rays is $<10\%$ of the \chandra\ zeroth-order ACIS-S observation, so the influence of emission from shocked gas along the wind--wind boundary is not very significant {, though contamination may be larger in some regions of the spectrum, depending on the emission-measure temperature distribution of the colliding-wind
X-rays compared to that of the X-rays arising from embedded wind shocks}.  We caution, however, that the model X-ray flux is dependent on the boundary condition imposed at the surface of Aa2, and so imposing a condition where the incoming wind from star Aa1 shocks more strongly (weakly) will increase (decrease) the amount of X-ray emission from the wind--star collision.

\section{Conclusions}
\label{sec:conc}

Delta Ori Aa is an X-ray bright, nearby, eclipsing binary and so offers the potential to directly probe the X-ray emitting gas distribution in the primary star's wind as the secondary star revolves through the primary's wind. Our \chandra\ program was designed to obtain high signal-to-noise and high spectral resolution spectrometry of this system throughout an entire orbit.  In this paper, we have sought to characterize the overall spectrum at its highest signal-to-noise ratio by combining all of the \chandra\ spectra and examining temperature distributions and line parameters. Our main results are presented below.

\begin{enumerate}
\item Our analysis of the \chandra\ image shows that the emission is mostly dominated by \dori a, with little detectable emission from $\delta$~Ori~Ab.

\item The temperature distribution of the X-ray emitting gas can be characterized by three dominant temperatures, which agrees fairly well with the temperature distributions derived by the earlier analysis of \citet{2002ApJ...577..951M} and \citet{Raassen:2013fk}.  

\item The strong lines are generally symmetric, and Gaussian profiles provide a reasonable representation of the profile shape, though in most cases, and especially for the \ion{Ne}{10} and \ion{Fe}{17} there are significant deviations from Gaussian symmetry.  

\item The line widths determined by Gaussian modeling shows that half-widths are typically $0.3-0.5\times V_{\infty}$, where $V_{\infty}$ is the terminal velocity of the wind of \dori a1.  These values are generally larger than the line widths measured by \cite{2002ApJ...577..951M}, though it is unclear whether this represents a real change in the line profile or if there is a calibration issue in the analysis of the earlier data set, which was obtained at an anomalously high focal plane temperature.

\item We find a strong  {anti-}correlation between the widths of the H-like and He-like transitions and the excitation energy.  {This indicates that the lower-energy transitions occur in a region
with larger velocities. Assuming a standard wind acceleration law, this correlation probably indicates that the lower-energy lines emerge from further out in the wind}.

\item Analysis of strong and weak transitions of \ion{Fe}{17} and \ion{Ne}{10} indicates that resonance scattering may be important in determining the flux and/or shape of the stronger line.  This agrees with the analysis of the \ion{Fe}{17} line by \cite{2002ApJ...577..951M} but at higher significance.     {We caution that some of these} differences  {in the observed to predicted line ratios may be influenced} by  {an inaccurate} temperature distribution and/or  {uncertainties in the atomic physics. It is also interesting to note that these two lines also have the most non-Gaussian profiles, as shown in Figure \ref{fig:h-like}, perhaps indicative that some line photons have been scattered out of the line of sight.}

\end{enumerate}

The spectrum combined from the four individual \chandra-HETGS observations represents a very high signal-to-noise view of the emission from \dori a.  However, these observations were obtained at a variety of orbital phases, so that the combined spectrum is a phase-averaged view of the overall X-ray emission from \dori a.  In a companion paper \citep{2015arXiv150704972N}  we look for the effects of phase- and time-dependent changes in the continuum and line spectrum.

\acknowledgments

We thank the MOST team for the award of observing time for \dori. We also thank our anonymous referee, whose comments significantly improved this paper. M.F.C. would like to thank John Houck and Michael Nowak for many helpful discussions concerning data analysis with ISIS. Support for this work was provided by the National Aeronautics and Space Administration through Chandra Award Number GO3-14015A and GO3-14015E issued by the Chandra X-ray Observatory Center, which is operated by the Smithsonian Astrophysical Observatory for and on behalf of the National Aeronautics Space Administration under contract NAS8-03060. M.F.C., J.S.N., W.L.W., C.M.P.R., and K.H. gratefully acknowledge this support.   M.F.C. acknowledges support from NASA under cooperative agreement number NNG06EO90A.  N.R.E. is grateful for support from the Chandra X-ray Center NASA Contract NAS8-03060. C.M.P.R. is supported by an appointment to the NASA Postdoctoral Program at the Goddard Space Flight Center, administered by Oak Ridge Associated Universities through a contract with NASA. T.S. is grateful for financial support from the Leibniz Graduate School for Quantitative Spectroscopy in Astrophysics, a joint project of the Leibniz Institute for Astrophysics Potsdam (AIP) and the institute of Physics and Astronomy of the University of Potsdam. Y.N. acknowledges support from the Fonds National de la Recherche Scientifique (Belgium), the Communaut\'e Fran\c caise de Belgique, the PRODEX \xmm\ and \integral\ contracts, and the `Action de Recherche Concert\'ee' (CFWB-Acad\'emie Wallonie Europe). N.D.R. gratefully acknowledges his CRAQ (Centre de Recherche en Astrophysique du Qu\'ebec) fellowship. A.F.J.M. is grateful for financial support from NSERC (Canada) and FRQNT (Quebec). J.L.H. acknowledges support from NASA award NNX13AF40G and NSF award AST-0807477. This research has made use of NASA's Astrophysics Data System. This research has made use of data and/or software provided by the High Energy Astrophysics Science Archive Research Center (HEASARC), which is a service of the Astrophysics Science Division at NASA/GSFC and the High Energy Astrophysics Division of the Smithsonian Astrophysical Observatory. This research made use of the Chandra Transmission Grating Catalog and archive (http://tgcat.mit.edu). The SPH simulations presented in this paper made use of the resources provided by the NASA High-End Computing (HEC) Program through the NASA Advanced Supercomputing (NAS) Division at Ames Research Center.

\bibliography{mfc_bibliography_dori}

\begin{thebibliography}{}
\expandafter\ifx\csname natexlab\endcsname\relax\def\natexlab#1{#1}\fi

\bibitem[{{Arnaud}(1996)}]{Arnaud96}
{Arnaud}, K.~A. 1996, in Astronomical Society of the Pacific Conference Series,
  Vol. 101, Astronomical Data Analysis Software and Systems V, ed. G.~H.
  {Jacoby} \& J.~{Barnes}, 17

\bibitem[{{Asplund} {et~al.}(2009){Asplund}, {Grevesse}, {Sauval}, \&
  {Scott}}]{AsplundP09}
{Asplund}, M., {Grevesse}, N., {Sauval}, A.~J., \& {Scott}, P. 2009, \araa, 47,
  481

\bibitem[{{Bate} {et~al.}(1995){Bate}, {Bonnell}, \&
  {Price}}]{BateBonnellPrice95}
{Bate}, M.~R., {Bonnell}, I.~A., \& {Price}, N.~M. 1995, \mnras, 277, 362

\bibitem[{{Benz}(1990)}]{Benz90}
{Benz}, W. 1990, in Numerical Modelling of Nonlinear Stellar Pulsations
  Problems and Prospects, ed. J.~R. {Buchler}, 269

\bibitem[{{Berghoefer} {et~al.}(1997){Berghoefer}, {Schmitt}, {Danner}, \&
  {Cassinelli}}]{1997A&A...322..167B}
{Berghoefer}, T.~W., {Schmitt}, J.~H.~M.~M., {Danner}, R., \& {Cassinelli},
  J.~P. 1997, \aap, 322, 167

\bibitem[{{Bernitt} {et~al.}(2012){Bernitt}, {Brown}, {Rudolph},
  {Steinbr{\"u}gge}, {Graf}, {Leutenegger}, {Epp}, {Eberle}, {Kubi{\v c}ek},
  {M{\"a}ckel}, {Simon}, {Tr{\"a}bert}, {Magee}, {Beilmann}, {Hell},
  {Schippers}, {M{\"u}ller}, {Kahn}, {Surzhykov}, {Harman}, {Keitel},
  {Clementson}, {Porter}, {Schlotter}, {Turner}, {Ullrich}, {Beiersdorfer}, \&
  {L{\'o}pez-Urrutia}}]{Bernitt:2012kx}
{Bernitt}, S., {Brown}, G.~V., {Rudolph}, J.~K., {et~al.} 2012, \nat, 492, 225

\bibitem[{{Brinkman} {et~al.}(1987){Brinkman}, {van Rooijen}, {Bleeker},
  {Dijkstra}, {Heise}, {de Korte}, {Mewe}, \& {Paerels}}]{Brinkman:1987rf}
{Brinkman}, A.~C., {van Rooijen}, J.~J., {Bleeker}, J.~A.~M., {et~al.} 1987,
  Astrophysical Letters and Communications, 26, 73

\bibitem[{{Caballero} \& {Solano}(2008)}]{Caballero:2008oq}
{Caballero}, J.~A., \& {Solano}, E. 2008, \aap, 485, 931

\bibitem[{{Canizares} {et~al.}(2005){Canizares}, {Davis}, {Dewey}, {Flanagan},
  {Galton}, {Huenemoerder}, {Ishibashi}, {Markert}, {Marshall}, {McGuirk},
  {Schattenburg}, {Schulz}, {Smith}, \& {Wise}}]{Canizares:2005bh}
{Canizares}, C.~R., {Davis}, J.~E., {Dewey}, D., {et~al.} 2005, \pasp, 117,
  1144

\bibitem[{{Cantiello} {et~al.}(2009){Cantiello}, {Langer}, {Brott}, {de Koter},
  {Shore}, {Vink}, {Voegler}, {Lennon}, \& {Yoon}}]{Cantiello:2009kq}
{Cantiello}, M., {Langer}, N., {Brott}, I., {et~al.} 2009, \aap, 499, 279

\bibitem[{{Cassinelli} \& {Swank}(1983)}]{Cassinelli:1983qe}
{Cassinelli}, J.~P., \& {Swank}, J.~H. 1983, \apj, 271, 681

\bibitem[{{Castor} {et~al.}(1975){Castor}, {Abbott}, \&
  {Klein}}]{CastorAbbottKlein75}
{Castor}, J.~I., {Abbott}, D.~C., \& {Klein}, R.~I. 1975, \apj, 195, 157

\bibitem[{{Chlebowski} {et~al.}(1989){Chlebowski}, {Harnden}, \&
  {Sciortino}}]{Chlebowski:1989cr}
{Chlebowski}, T., {Harnden}, Jr., F.~R., \& {Sciortino}, S. 1989, \apj, 341,
  427

\bibitem[{{Cohen} {et~al.}(2014){Cohen}, {Wollman}, {Leutenegger}, {Sundqvist},
  {Fullerton}, {Zsargo}, \& {Owocki}}]{Cohen:2014lr}
{Cohen}, D.~H., {Wollman}, E.~E., {Leutenegger}, M.~A., {et~al.} 2014, ArXiv
  e-prints, arXiv:1401.7995

\bibitem[{{Corcoran} {et~al.}(1994){Corcoran}, {Waldron}, {Macfarlane}, {Chen},
  {Pollock}, {Torii}, {Kitamoto}, {Miura}, {Egoshi}, \&
  {Ohno}}]{1994ApJ...436L..95C}
{Corcoran}, M.~F., {Waldron}, W.~L., {Macfarlane}, J.~J., {et~al.} 1994, \apjl,
  436, L95

\bibitem[{{Corcoran} {et~al.}(2015){Corcoran}, {Nichols}, {Pablo}, {Shenar},
  {Pollock}, {Waldron}, {Moffat}, {Richardson}, {Russell}, {Hamaguchi},
  {Huenemoerder}, {Oskinova}, {Hamann}, {Naze}, {Ignace}, {Evans}, {Lomax},
  {Hoffman}, {Gayley}, {Owocki}, {Leutenegger}, {Gull}, {Hole}, {Lauer}, \&
  {Iping}}]{2015arXiv150705101C}
{Corcoran}, M.~F., {Nichols}, J.~S., {Pablo}, H., {et~al.} 2015, ArXiv
  e-prints, arXiv:1507.05101

\bibitem[{{Fisher} \& {Meyerott}(1964)}]{Fisher:1964ij}
{Fisher}, P.~C., \& {Meyerott}, A.~J. 1964, \apj, 139, 123

\bibitem[{{Foster} {et~al.}(2012){Foster}, {Ji}, {Smith}, \&
  {Brickhouse}}]{Foster:2012sf}
{Foster}, A.~R., {Ji}, L., {Smith}, R.~K., \& {Brickhouse}, N.~S. 2012, \apj,
  756, 128

\bibitem[{{Friend} \& {Abbott}(1986)}]{FriendAbbott86}
{Friend}, D.~B., \& {Abbott}, D.~C. 1986, \apj, 311, 701

\bibitem[{{Fruscione} {et~al.}(2006){Fruscione}, {McDowell}, {Allen},
  {Brickhouse}, {Burke}, {Davis}, {Durham}, {Elvis}, {Galle}, {Harris},
  {Huenemoerder}, {Houck}, {Ishibashi}, {Karovska}, {Nicastro}, {Noble},
  {Nowak}, {Primini}, {Siemiginowska}, {Smith}, \& {Wise}}]{Fruscione:2006uq}
{Fruscione}, A., {McDowell}, J.~C., {Allen}, G.~E., {et~al.} 2006, in Society
  of Photo-Optical Instrumentation Engineers (SPIE) Conference Series, Vol.
  6270, Society of Photo-Optical Instrumentation Engineers (SPIE) Conference
  Series, 1

\bibitem[{{Gayley}(1995)}]{Gayley95}
{Gayley}, K.~G. 1995, \apj, 454, 410

\bibitem[{{Gayley} {et~al.}(1997){Gayley}, {Owocki}, \&
  {Cranmer}}]{GayleyOwockiCranmer97}
{Gayley}, K.~G., {Owocki}, S.~P., \& {Cranmer}, S.~R. 1997, \apj, 475, 786

\bibitem[{{Gr{\"a}fener} {et~al.}(2002){Gr{\"a}fener}, {Koesterke}, \&
  {Hamann}}]{Grafener:2002vn}
{Gr{\"a}fener}, G., {Koesterke}, L., \& {Hamann}, W.-R. 2002, \aap, 387, 244

\bibitem[{{Grant} {et~al.}(2006){Grant}, {Bautz}, {Kissel}, {LaMarr}, \&
  {Prigozhin}}]{Grant:2006kq}
{Grant}, C.~E., {Bautz}, M.~W., {Kissel}, S.~E., {LaMarr}, B., \& {Prigozhin},
  G.~Y. 2006, in Society of Photo-Optical Instrumentation Engineers (SPIE)
  Conference Series, Vol. 6276, Society of Photo-Optical Instrumentation
  Engineers (SPIE) Conference Series, 1

\bibitem[{{Haberl} \& {White}(1993)}]{1993A&A...280..519H}
{Haberl}, F., \& {White}, N.~E. 1993, \aap, 280, 519

\bibitem[{{Hamann} \& {Gr{\"a}fener}(2003)}]{Hamann:2003rt}
{Hamann}, W.-R., \& {Gr{\"a}fener}, G. 2003, \aap, 410, 993

\bibitem[{{Harmanec} {et~al.}(2013){Harmanec}, {Mayer}, \& {{\v
  S}lechta}}]{Harmanec:2013lr}
{Harmanec}, P., {Mayer}, P., \& {{\v S}lechta}, M. 2013, in Massive Stars: From
  Alpha to Omega, 70

\bibitem[{{Hartmann}(1904)}]{Hartmann:1904ly}
{Hartmann}, J. 1904, \apj, 19, 268

\bibitem[{{Harvin} {et~al.}(2002){Harvin}, {Gies}, {Bagnuolo}, {Penny}, \&
  {Thaller}}]{2002ApJ...565.1216H}
{Harvin}, J.~A., {Gies}, D.~R., {Bagnuolo}, Jr., W.~G., {Penny}, L.~R., \&
  {Thaller}, M.~L. 2002, \apj, 565, 1216

\bibitem[{{Herv{\'e}} {et~al.}(2013){Herv{\'e}}, {Rauw}, \&
  {Naz{\'e}}}]{Herve:2013rt}
{Herv{\'e}}, A., {Rauw}, G., \& {Naz{\'e}}, Y. 2013, \aap, 551, A83

\bibitem[{{Houck} \& {Denicola}(2000)}]{Houck:2000qd}
{Houck}, J.~C., \& {Denicola}, L.~A. 2000, in Astronomical Society of the
  Pacific Conference Series, Vol. 216, Astronomical Data Analysis Software and
  Systems IX, ed. N.~{Manset}, C.~{Veillet}, \& D.~{Crabtree}, 591

\bibitem[{{Huenemoerder} {et~al.}(2011){Huenemoerder}, {Mitschang}, {Dewey},
  {Nowak}, {Schulz}, {Nichols}, {Davis}, {Houck}, {Marshall}, {Noble},
  {Morgan}, \& {Canizares}}]{Huenemoerder:2011vn}
{Huenemoerder}, D.~P., {Mitschang}, A., {Dewey}, D., {et~al.} 2011, \aj, 141,
  129

\bibitem[{{Koch} \& {Hrivnak}(1981)}]{Koch:1981mz}
{Koch}, R.~H., \& {Hrivnak}, B.~J. 1981, \apj, 248, 249

\bibitem[{{Kudritzki} \& {Puls}(2000)}]{2000ARA&A..38..613K}
{Kudritzki}, R., \& {Puls}, J. 2000, \araa, 38, 613

\bibitem[{{Leutenegger} {et~al.}(2010){Leutenegger}, {Cohen}, {Zsarg{\'o}},
  {Martell}, {MacArthur}, {Owocki}, {Gagn{\'e}}, \& {Hillier}}]{LeuteneggerP10}
{Leutenegger}, M.~A., {Cohen}, D.~H., {Zsarg{\'o}}, J., {et~al.} 2010, \apj,
  719, 1767

\bibitem[{{Leutenegger} {et~al.}(2006){Leutenegger}, {Paerels}, {Kahn}, \&
  {Cohen}}]{2006ApJ...650.1096L}
{Leutenegger}, M.~A., {Paerels}, F.~B.~S., {Kahn}, S.~M., \& {Cohen}, D.~H.
  2006, \apj, 650, 1096

\bibitem[{{Long} \& {White}(1980)}]{Long:1980dp}
{Long}, K.~S., \& {White}, R.~L. 1980, \apjl, 239, L65

\bibitem[{{Lucy} \& {White}(1980)}]{1980ApJ...241..300L}
{Lucy}, L.~B., \& {White}, R.~L. 1980, \apj, 241, 300

\bibitem[{{Madura} {et~al.}(2013){Madura}, {Gull}, {Okazaki}, {Russell},
  {Owocki}, {Groh}, {Corcoran}, {Hamaguchi}, \& {Teodoro}}]{MaduraP13}
{Madura}, T.~I., {Gull}, T.~R., {Okazaki}, A.~T., {et~al.} 2013, \mnras, 436,
  3820

\bibitem[{{Ma{\'{\i}}z Apell{\'a}niz} {et~al.}(2013){Ma{\'{\i}}z
  Apell{\'a}niz}, {Sota}, {Morrell}, {Barb{\'a}}, {Walborn}, {Alfaro}, {Gamen},
  {Arias}, \& {Gallego Calvente}}]{Maiz-Apellaniz:2013uq}
{Ma{\'{\i}}z Apell{\'a}niz}, J., {Sota}, A., {Morrell}, N.~I., {et~al.} 2013,
  in Massive Stars: From alpha to Omega, 198

\bibitem[{{Mayer} {et~al.}(2010){Mayer}, {Harmanec}, {Wolf}, {Bo{\v z}i{\'c}},
  \& {{\v S}lechta}}]{2010A&A...520A..89M}
{Mayer}, P., {Harmanec}, P., {Wolf}, M., {Bo{\v z}i{\'c}}, H., \& {{\v
  S}lechta}, M. 2010, \aap, 520, A89+

\bibitem[{{Miller} {et~al.}(2002){Miller}, {Cassinelli}, {Waldron},
  {MacFarlane}, \& {Cohen}}]{2002ApJ...577..951M}
{Miller}, N.~A., {Cassinelli}, J.~P., {Waldron}, W.~L., {MacFarlane}, J.~J., \&
  {Cohen}, D.~H. 2002, \apj, 577, 951

\bibitem[{{Nichols} {et~al.}(2015){Nichols}, {Huenemoerder}, {Corcoran},
  {Waldron}, {Naz{\'e}}, {Pollock}, {Moffat}, {Lauer}, {Shenar}, {Russell},
  {Richardson}, {Pablo}, {Evans}, {Hamaguchi}, {Gull}, {Hamann}, {Oskinova},
  {Ignace}, {Hoffman}, {Hole}, \& {Lomax}}]{2015arXiv150704972N}
{Nichols}, J.~S., {Huenemoerder}, D.~P., {Corcoran}, M.~F., {et~al.} 2015,
  ArXiv e-prints, arXiv:1507.04972

\bibitem[{{Okazaki} {et~al.}(2008){Okazaki}, {Owocki}, {Russell}, \&
  {Corcoran}}]{OkazakiP08}
{Okazaki}, A.~T., {Owocki}, S.~P., {Russell}, C.~M.~P., \& {Corcoran}, M.~F.
  2008, \mnras, 388, L39

\bibitem[{{Oskinova} {et~al.}(2006){Oskinova}, {Feldmeier}, \&
  {Hamann}}]{Oskinova:2006dk}
{Oskinova}, L.~M., {Feldmeier}, A., \& {Hamann}, W.-R. 2006, \mnras, 372, 313

\bibitem[{{Owocki} \& {Cohen}(2006)}]{Owocki:2006fr}
{Owocki}, S.~P., \& {Cohen}, D.~H. 2006, \apj, 648, 565

\bibitem[{{Owocki} {et~al.}(1996){Owocki}, {Cranmer}, \&
  {Gayley}}]{1996ApJ...472L.115O}
{Owocki}, S.~P., {Cranmer}, S.~R., \& {Gayley}, K.~G. 1996, \apjl, 472, L115+

\bibitem[{{Owocki} \& {Gayley}(1995)}]{OwockiGayley95}
{Owocki}, S.~P., \& {Gayley}, K.~G. 1995, \apjl, 454, L145

\bibitem[{{Pablo} {et~al.}(2015){Pablo}, {Richardson}, {Moffat}, {Corcoran},
  {Shenar}, {Benvenuto}, {Fuller}, {Naze}, {Hoffman}, {Miroshnichenko}, {Maiz
  Apellaniz}, {Evans}, {Eversberg}, {Gayley}, {Gull}, {Hamaguch}, {Hamann},
  {Henrichs}, {Hole}, {Ignace}, {Iping}, {Lauer}, {Leutenegger}, {Lomax},
  {Nichols}, {Oskinova}, {Owocki}, {Pollock}, {Russell}, {Waldron}, {Buil},
  {Garrel}, {Graham}, {Heathcote}, {Lemoult}, {Li}, {Mauclaire}, {Potter},
  {Ribeiro}, {Matthews}, {Cameron}, {Guenther}, {Kuschnig}, {Rowe}, {Rucinski},
  {Sasselov}, \& {Weiss}}]{2015arXiv150408002P}
{Pablo}, H., {Richardson}, N.~D., {Moffat}, A.~F.~J., {et~al.} 2015, ArXiv
  e-prints, arXiv:1504.08002

\bibitem[{{Pallavicini} {et~al.}(1981){Pallavicini}, {Golub}, {Rosner},
  {Vaiana}, {Ayres}, \& {Linsky}}]{Pallavicini:1981ai}
{Pallavicini}, R., {Golub}, L., {Rosner}, R., {et~al.} 1981, \apj, 248, 279

\bibitem[{{Pauldrach} {et~al.}(1986){Pauldrach}, {Puls}, \&
  {Kudritzki}}]{PauldrachPulsKudritzki86}
{Pauldrach}, A., {Puls}, J., \& {Kudritzki}, R.~P. 1986, \aap, 164, 86

\bibitem[{{Price}(2007)}]{Price07}
{Price}, D.~J. 2007, \pasa, 24, 159

\bibitem[{{Raassen} \& {Pollock}(2013)}]{Raassen:2013fk}
{Raassen}, A.~J.~J., \& {Pollock}, A.~M.~T. 2013, \aap, 550, A55

\bibitem[{{Richardson} {et~al.}(2015){Richardson}, {Moffat}, {Gull}, {Lindler},
  {Gies}, {Corcoran}, \& {Chen{\'e}}}]{Richardson:2015fk}
{Richardson}, N.~D., {Moffat}, A.~F.~J., {Gull}, T.~R., {et~al.} 2015, ArXiv
  e-prints, arXiv:1506.05530

\bibitem[{{Russell}(2013)}]{Russell13}
{Russell}, C.~M.~P. 2013, PhD thesis, University of Delaware

\bibitem[{{Shenar} {et~al.}(2015){Shenar}, {Oskinova}, {Hamann}, {Corcoran},
  {Moffat}, {Pablo}, {Richardson}, {Waldron}, {Huenemoerder}, {Ma{\'{\i}}z
  Apell{\'a}niz}, {Nichols}, {Todt}, {Naz{\'e}}, {Hoffman}, {Pollock}, \&
  {Negueruela}}]{2015arXiv150303476S}
{Shenar}, T., {Oskinova}, L., {Hamann}, W.-R., {et~al.} 2015, ArXiv e-prints,
  arXiv:1503.03476

\bibitem[{{Smith} \& {Brickhouse}(2000)}]{Smith:2000qv}
{Smith}, R.~K., \& {Brickhouse}, N.~S. 2000, in Revista Mexicana de Astronomia
  y Astrofisica Conference Series, Vol.~9, Revista Mexicana de Astronomia y
  Astrofisica Conference Series, ed. S.~J. {Arthur}, N.~S. {Brickhouse}, \&
  J.~{Franco}, 134--136

\bibitem[{{Smith} {et~al.}(2001){Smith}, {Brickhouse}, {Liedahl}, \&
  {Raymond}}]{SmithP01}
{Smith}, R.~K., {Brickhouse}, N.~S., {Liedahl}, D.~A., \& {Raymond}, J.~C.
  2001, \apjl, 556, L91

\bibitem[{{Snow} {et~al.}(1981){Snow}, {Cash}, \& {Grady}}]{Snow:1981tg}
{Snow}, Jr., T.~P., {Cash}, W., \& {Grady}, C.~A. 1981, \apjl, 244, L19

\bibitem[{{Stebbins}(1915)}]{Stebbins:1915gf}
{Stebbins}, J. 1915, \apj, 42, 133

\bibitem[{{Stevens} {et~al.}(1992){Stevens}, {Blondin}, \&
  {Pollock}}]{StevensBlondinPollock92}
{Stevens}, I.~R., {Blondin}, J.~M., \& {Pollock}, A.~M.~T. 1992, \apj, 386, 265

\bibitem[{{Stevens} \& {Pollock}(1994)}]{StevensPollock94}
{Stevens}, I.~R., \& {Pollock}, A.~M.~T. 1994, \mnras, 269, 226

\bibitem[{{Tokovinin} {et~al.}(2014){Tokovinin}, {Mason}, \&
  {Hartkopf}}]{Tokovinin:2014tg}
{Tokovinin}, A., {Mason}, B.~D., \& {Hartkopf}, W.~I. 2014, \aj, 147, 123

\bibitem[{{Townsend}(2009)}]{Townsend09}
{Townsend}, R.~H.~D. 2009, \apjs, 181, 391

\bibitem[{{Walborn} {et~al.}(2009){Walborn}, {Nichols}, \&
  {Waldron}}]{2009ApJ...703..633W}
{Walborn}, N.~R., {Nichols}, J.~S., \& {Waldron}, W.~L. 2009, \apj, 703, 633

\bibitem[{{Waldron} \& {Cassinelli}(2001)}]{Waldron:2001lr}
{Waldron}, W.~L., \& {Cassinelli}, J.~P. 2001, \apjl, 548, L45

\bibitem[{{Waldron} \& {Cassinelli}(2007)}]{2007ApJ...668..456W}
---. 2007, \apj, 668, 456

\bibitem[{{Walker} {et~al.}(2003){Walker}, {Matthews}, {Kuschnig}, {Johnson},
  {Rucinski}, {Pazder}, {Burley}, {Walker}, {Skaret}, {Zee}, {Grocott},
  {Carroll}, {Sinclair}, {Sturgeon}, \& {Harron}}]{Walker:2003yq}
{Walker}, G., {Matthews}, J., {Kuschnig}, R., {et~al.} 2003, \pasp, 115, 1023

\bibitem[{{Wilms} {et~al.}(2000{\natexlab{a}}){Wilms}, {Allen}, \&
  {McCray}}]{Wilms:2000rr}
{Wilms}, J., {Allen}, A., \& {McCray}, R. 2000{\natexlab{a}}, \apj, 542, 914

\bibitem[{{Wilms} {et~al.}(2000{\natexlab{b}}){Wilms}, {Allen}, \&
  {McCray}}]{WilmsAllenMcCray00}
---. 2000{\natexlab{b}}, \apj, 542, 914

\end{thebibliography}

\end{document}